\documentclass[twocolumn,dvipsnames]{aastex631}
\usepackage{subfigure}

\newcommand\msun{\hbox{${ M}_{\odot}$}}
\newcommand\mstar{\hbox{${ M}_{\star}$}}

\received{October 16, 2023}
\revised{February 13, 2024}
\accepted{February 19, 2024}

\shorttitle{MW Analogues}
\shortauthors{Tan et al.}
\graphicspath{{./}{}}

\begin{document}

\title{A Measurement of the Assembly of Milky Way Analogues at Redshifts $0.5 < z < 2$ with Resolved Stellar Mass and Star-Formation Rate Profiles}

\author[0000-0002-3503-8899]{Vivian Yun Yan Tan}
\affiliation{Department of Physics and Astronomy, York University, 4700 Keele Street, Toronto, ON, M3J 1P3, Canada}
\author[0000-0002-9330-9108]{Adam Muzzin}
\affiliation{Department of Physics and Astronomy, York University, 4700 Keele Street, Toronto, ON, M3J 1P3, Canada}
\author[0000-0001-9002-3502]{Danilo Marchesini}
\affiliation{Department of Physics and Astronomy, Tufts University, 574 Boston Avenue, MA 02155, USA}
\author[0000-0003-0780-9526]{Visal Sok}
\affiliation{Department of Physics and Astronomy, York University, 4700 Keele Street, Toronto, ON, M3J 1P3, Canada}
\author[0000-0001-8830-2166]{Ghassan T E Sarrouh}
\affiliation{Department of Physics and Astronomy, York University, 4700 Keele Street, Toronto, ON, M3J 1P3, Canada}
\author[0000-0002-7248-1566]{Z. Cemile Marsan}
\affiliation{Department of Physics and Astronomy, York University, 4700 Keele Street, Toronto, ON, M3J 1P3, Canada}


\begin{abstract}
\noindent The resolved mass assembly of Milky-Way-mass galaxies has been previously studied in simulations, the local universe, and at higher redshifts using infrared (IR) light profiles.  To better characterize the mass assembly of Milky Way Analogues (MWAs), as well as their changes in star-formation rate and color gradients, we construct resolved stellar mass and star-formation rate maps of MWA progenitors selected with abundance matching techniques up to z $\sim$ 2 using deep, multi-wavelength imaging data from the Hubble Frontier Fields.  Our results using stellar mass profiles agree well with previous studies that utilize IR light profiles, showing that the inner 2 kpc of the galaxies and the regions beyond 2 kpc exhibit similar rates of stellar mass growth. This indicates the progenitors of MWAs from $z\sim 2$ to the present do not preferentially grow their bulges or their disks. The evolution of the star-formation rate (SFR) profiles indicate greater decrease in SFR density in the inner regions versus the outer regions. S\'ersic parameters indicate modest growth in the central regions at lower redshifts, perhaps indicating slight bulge growth. However, the S\'ersic index does not rise above $n \sim 2$ until $z < 0.5$, meaning these galaxies are still disk dominated systems. We find that the half-mass radii of the MWA progenitors increase between $1.5 < z < 2$, but remain constant at later epochs ($z < 1.5$). This implies mild bulge growth since $z\sim 2$ in MWA progenitors, in line with previous MWA mass assembly studies. 
\end{abstract}


\section{Introduction}\label{sec:intro}
Observations of stars and gas within the Milky Way have long been the primary method of deciphering its formation and evolutionary history (\citealt{Freeman:2002, Freeman:2013, Bovy:2012,Bovy:2013,Rix:2013}, see \citealt{Bland-Hawthorn:2016,Helmi:2020} for reviews). However, the Milky Way is certainly not unique, and is a rather typical galaxy in the universe. It is a spiral galaxy with a mass of $5-6\times 10^{10} \msun$ \citep{Mutch:2011, Licquia:2015b, Bland-Hawthorn:2016} which exists in a low density environment, and it currently resides in the green valley, the transition state between star-forming and quiescent. It has a bulge with an extent of $R \sim 2$~kpc in the major axis, a bar with half-length $\sim 5$kpc \citep{Wegg:2015}, and a stellar disk that extends out to $R \sim 15$kpc \citep{Robin:1992,Ruphy:1996,Minniti:2011}. 
We wish to place the Milky Way in the context of how it formed, especially in relation to Milky Way Analogues in the local and the higher redshift universe. One way to do this is to utilize observations of Milky Way Analogues (MWAs) at higher redshifts to trace the evolution of our galaxy over cosmic time. There are many ways to define MWAs; however, for the purpose of this paper we define them only by stellar mass.

The morphology of star-forming galaxies in the local universe suggests that star formation occurred in two phases, with bulges forming first and disks forming around them later, also known as ``inside-out" formation \citep{vandenBosch:1998, Munoz-Mateos:2007,Wang:2011, Goddard:2017}, with even the Milky Way's inner and outer disk forming in phases as well \citep{Haywood:2013}. An alternative picture posits that the stellar disks formed first, and then bulges grew later through secular processes \citep{Dekel:2009,Martig:2009}, which may be the case for lower mass galaxies but not higher mass ones, which tend to quench inside-out \citep{Cappellari:2013}. However, observations in the local universe show dwarf galaxies may have had alternating periods of outside-in and inside-out growth \citep{Ibarra-Medel:2016}. There is also evidence that outside-in quenching may be linked to belonging to denser environments such as clusters and large groups \citep{Gavazzi:2013,Schaefer:2017}. It is still an open question whether galaxies like the Milky Way formed inside-out or outside-in, or if the color gradient present in the Milky Way and similar spirals developed in a different way.

Specifically for MWAs, some studies have approached understanding the formation of MWAs from simulations \citep{Abadi:2003a,Abadi:2003b,Onorbe:2015,Torrey:2017,Buck:2020,Sotillo-Ramos:2022,Hasheminia:2022}. \cite{Buck:2020} find MW-like stellar disks form most of their mass in-situ and only 5\% of the mass is from mergers. \cite{Sotillo-Ramos:2022} find that MWAs can even survive major mergers with their disk intact but somewhat thicker. \cite{Hasheminia:2022} found MWAs have no evolution in their half-mass radii for the last 10 Gyr, which coincides with redshift $z\sim2$. While simulations offer one solution to uncovering a dynamic picture of the formation history of the Milky Way, there are limitations to the utility of their results. This is related to the particle resolution, the size of the region being simulated, the treatment of gas, stars, and dark matter, and other factors such as cosmological parameters or hydrodynamics, which have different treatments depending on which simulation code is used.

 There are also abundance matching studies that focus on the progenitors of massive galaxies in the local universe ($\star > 10^{11} \msun$, such as \cite{Ownsworth:2014, Mundy:2015, Ownsworth:2016}, and \cite{Hill:2017a}. \cite{Ownsworth:2014} found that major and minor mergers contribute up to 17\% and 34\% respectively of added stellar mass seen at $z=0.3$ since $z\sim3$, in comparison to the 24\% added by star-formation, by comparing merger-adjusted number densities to constant number densities. However, massive galaxies are more likely to be quiescent, and are on average less star-forming compared to the Milky Way. \cite{Mundy:2015} is a simulation-based study which found that the choice of number density in past epochs affects the recovered SFR and SFH of progenitors. \cite{Ownsworth:2016} and \cite{Hill:2017a} both find the quiescent fraction to be $\sim90\%$ at $z < 1$, but the former used a fixed cumulative number density and the latter used an evolving cumulative number density. Therefore, at $1 <z < 2$, \cite{Ownsworth:2016} find the majority of their massive galaxy progenitors to be quiescent, while for the progenitors in \cite{Hill:2017a}, they are still majority star-forming at the same redshift. Nevertheless, both studies agree that at earlier epochs ($z \gtrsim 2$), massive galaxies assembled mass primarily via star-formation, while at later epochs, most mass was assembled via mergers.

Three studies that use an observational approach, and which are seminal to furthering our understanding of the evolution of MWAs up to high redshift are \cite{vanDokkum:2013}, \cite{Patel:2013}, and \cite{Papovich:2015}.
 \cite{vanDokkum:2013} (hereafter VD13) used a constant number density through time to select MWAs at redshifts up to $z\sim 3$, with stellar mass functions from \cite{Marchesini:2009}. Their results show that the bulges and disks grow together in lockstep throughout the probed redshift. Using a constant comoving number density may not be the most accurate assumption to make for high redshift observations, due to galaxy mergers changing local density \citep{Behroozi:2013a,Behroozi:2013b,Leja:2013,Lin:2013}. 

 \cite{Patel:2013} (hereafter P13) used a different approach for choosing their sample: the star-forming main sequence. They used the relationship between specific star-formation rate (sSFR) to stellar mass, and chose their sample of MW progenitors from the sSFR-mass plane from \cite{Karim:2011}. The evolution of star-forming galaxies that led to a final mass was at $10^{10.5}\msun$, as computed by \cite{Leitner:2012}, was the basis of the stellar mass selection of their MWAs at higher redshift. On average, P13 find MWA progenitors to be more star-forming and less massive when compared to VD13 across the same redshift range. Interestingly, both studies agree in their findings that there are no strong indications for inside-out or outside-in growth in their progenitors, despite the different mass and redshift ranges. In fact, both works show that the inner and outer regions grow in lockstep with each other from $z \sim 3$ to the present epoch. 

\cite{Papovich:2015} (hereafter P15) utilized abundance matching in concordance with number density evaluations on the zFOURGE/CANDELS fields to define and delineate the different evolutionary phases that progenitors of MW-like and M31-like galaxies go through since $z \sim 3$. For the MW-like progenitors, the galaxies go from blue and star-forming at $z > 2$ to IR-bright and dust obscured from $1 < z < 1.8$, where the overall maximum SFR is measured, and SFR begins to steadily decrease as S\'ersic indices increase at $ z < 1$ to the present epoch. P15 also find that M31-like galaxies begin and end each evolutionary stage earlier than MW-like galaxies. All three of these studies listed above rely \textit{solely} on the observed light profiles to make arguments about morphology and mass assembly. Here we aim to extend the study of mass assembly of the MW by using SED-fitting to create spatially-resolved stellar mass profiles, as well as resolved star-formation rate profiles. This should bring a clearer and more accurate picture of the underlying morphological changes that progenitors of Milky Way-like galaxies undergo in the past 10 Gyr.

In this paper, we utilize abundance matching of galaxies from redshifts $0.5 < z < 2$ to select Milky Way analogues. We build upon the stellar mass map studies in \cite{Tan:2022}, from which we utilize the same methods to create resolved stellar mass and SFR profiles to trace the mass assembly for MW-like progenitors over time. Our abundance matching is based on the assumption that the approximate rank order of galaxies remains consistent throughout cosmic time, even as they grow in stellar mass, but also includes merger trees to account for slowly declining number density as galaxy haloes merge together due to gravitational interactions. Abundance matching techniques have previously been employed to study both MWAs and massive galaxies \citep{Yang:2012,Moster:2013,Marchesini:2014,Torrey:2017,Kravtsov:2018}. 

In \S \ref{sec:data}, we give further information on the Hubble Frontier Fields DeepSpace catalogs and which photometric filters were used, as well as the COSMOS/UltraVISTA survey from which the Schechter functions for comoving number densities is derived from. Our methods for constructing resolved 1-D and 2-D stellar mass and SFR profiles are outlined in detail in \S \ref{sec:methods}.  In \S \ref{sec:results}, we present the results of our analysis on the mass assembly and change in SFR over cosmic time for our sample of MWA progenitors. Our discussions of the implications of these results in comparison to previous works that used IR light profiles, as well as comparisons to simulations of MW progenitors are in \S \ref{sec:discussion}. For this work we assume a $\Lambda$CDM cosmological model of the universe with $\Omega_\Lambda$ = 0.7 and $\Omega_M$ = 0.3, along with a Hubble constant of $H_0 = 70$ km/s/Mpc. We assume a Chabrier IMF \citep{Chabrier:2003} for our determination of the stellar masses of our sample. Our magnitudes are reported in AB magnitudes.

\section{Data Sample} \label{sec:data}

\subsection{Hubble Frontier Fields}\label{sec:data1}

\begin{figure*}
    \centering
    \begin{subfigure}
        \centering
        \includegraphics[width=0.49\textwidth]{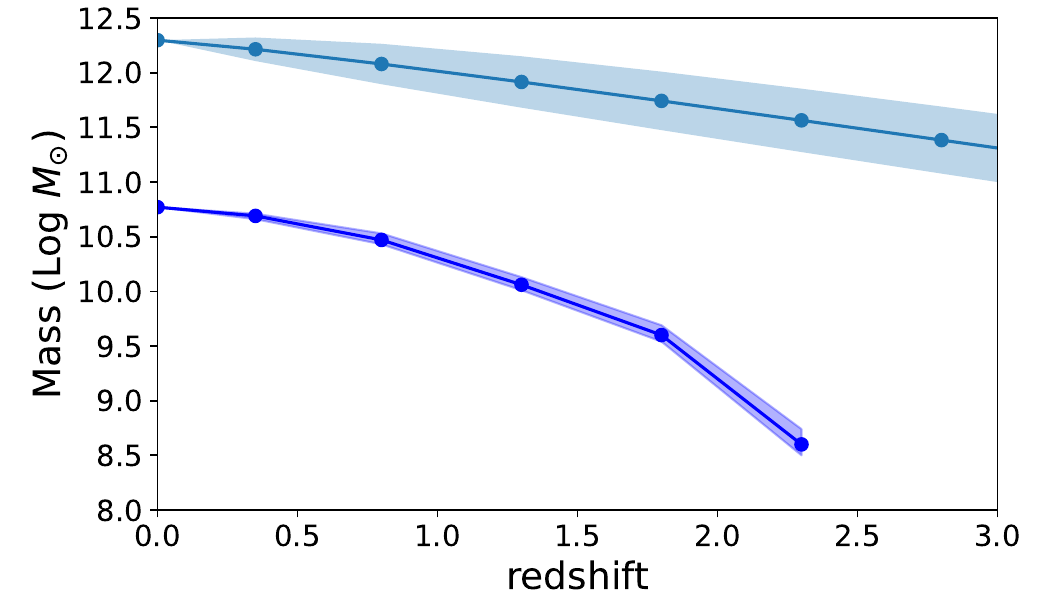}
    \end{subfigure}
    \begin{subfigure}
        \centering
        \includegraphics[width=0.49\textwidth]{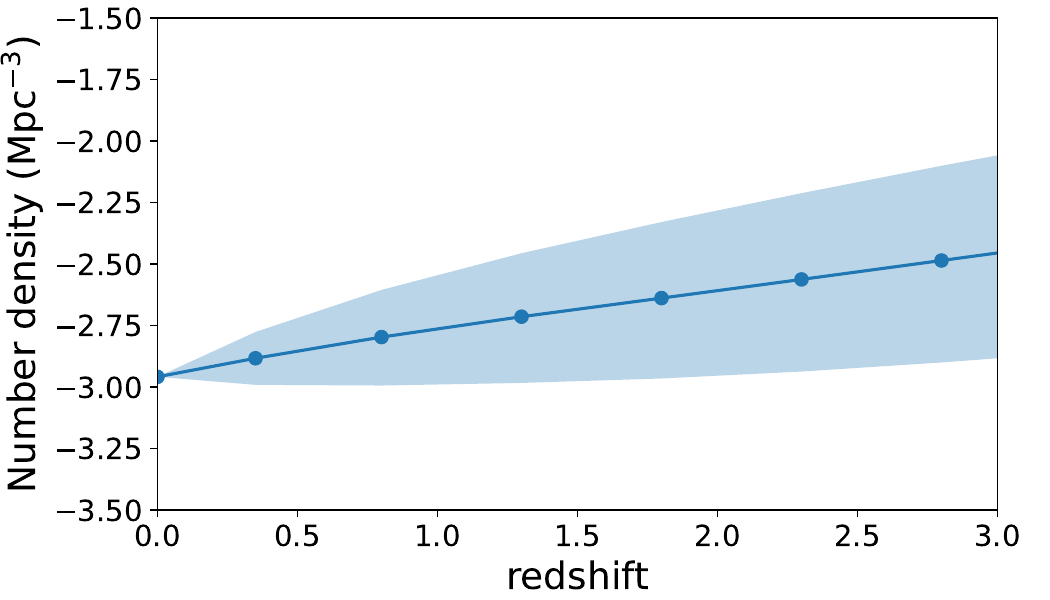}
    \end{subfigure}
    \caption{\emph{Left panel: }The evolution of the median halo mass (light blue) and the median stellar mass (dark blue) of MWAs from $z=3$ to $z=0$. \emph{Right panel:} The median integrated comoving number density for MWAs. The values at each redshift step is generated from the code by \cite{Behroozi:2013b}. The shaded regions represent 1$\sigma$ from the median. }\label{fig:num_density}
\end{figure*}

The Hubble Frontier Fields (HFF; \citealt{Lotz:2017}) are comprised of six lensing clusters at redshifts of $0.3 < z < 0.6$ and six flanking fields imaged in the UV, optical, and NIR with the Advanced Camera for Surveys (ACS) and Wide Field Camera 3 (WFC3) on the Hubble Space Telescope (HST).  We use the DeepSpace catalog from \cite{Shipley:2018} to select our sample of MWA progenitors. For this study, we primarily use the parallel flanking fields of the DeepSpace catalog, but also include the cluster fields as long as the objects' redshifts are outside the redshift range of the respective cluster's redshift by $z \pm 0.1$. This is because the Milky Way does not live within a cluster, so the number density of the fields is better suited to search for MWAs.


The redshift of our MWA sample spans from $0.3 < z < 2$, with only two galaxies at $z < 0.5$ (one at $z \sim 0.31$, another at $z\sim 0.42$). Based on the WFC3 and ACS PSF size, this implies an angular resolution of $0.25 - 0.50$ kpc per pixel and $0.8 - 1.5$. kpc per PSF FWHM.  In all modeling we use the images that are PSF-matched and convolved to the WFC3/F160W resolution from \cite{Shipley:2018} to ensure consistent resolution in each filter. The F160W filter has a PSF FWHM of $\sim0.177"$, which translates to a pixel size of 2.95 pixels. The PSFs for each filter are derived from stacking at least 3 or more isolated and unsaturated stars. The total science area covered by HFF DeepSpace is 156.1 arcmin$^2$. Although because we were not able to make use of Abell1063clu, the total area that is used in this work is only 150.5 arcmin$^2$.

For the flanking fields, each of the pointings has 7 filters (F435W, F606W, F814W, F105W, F125W, F140W, and F160W), except for the parallel field of MACS0416, which was imaged with 9 filters (the same 7 listed before, as well as F775W, and F850LP). For a more in-depth discussion on the DeepSpace catalog, we refer the reader to Sections 2 and 3 of \cite{Shipley:2018}. 

\subsection{Stellar Mass functions}


\begin{figure}
    \centering
    \includegraphics[width=\columnwidth]{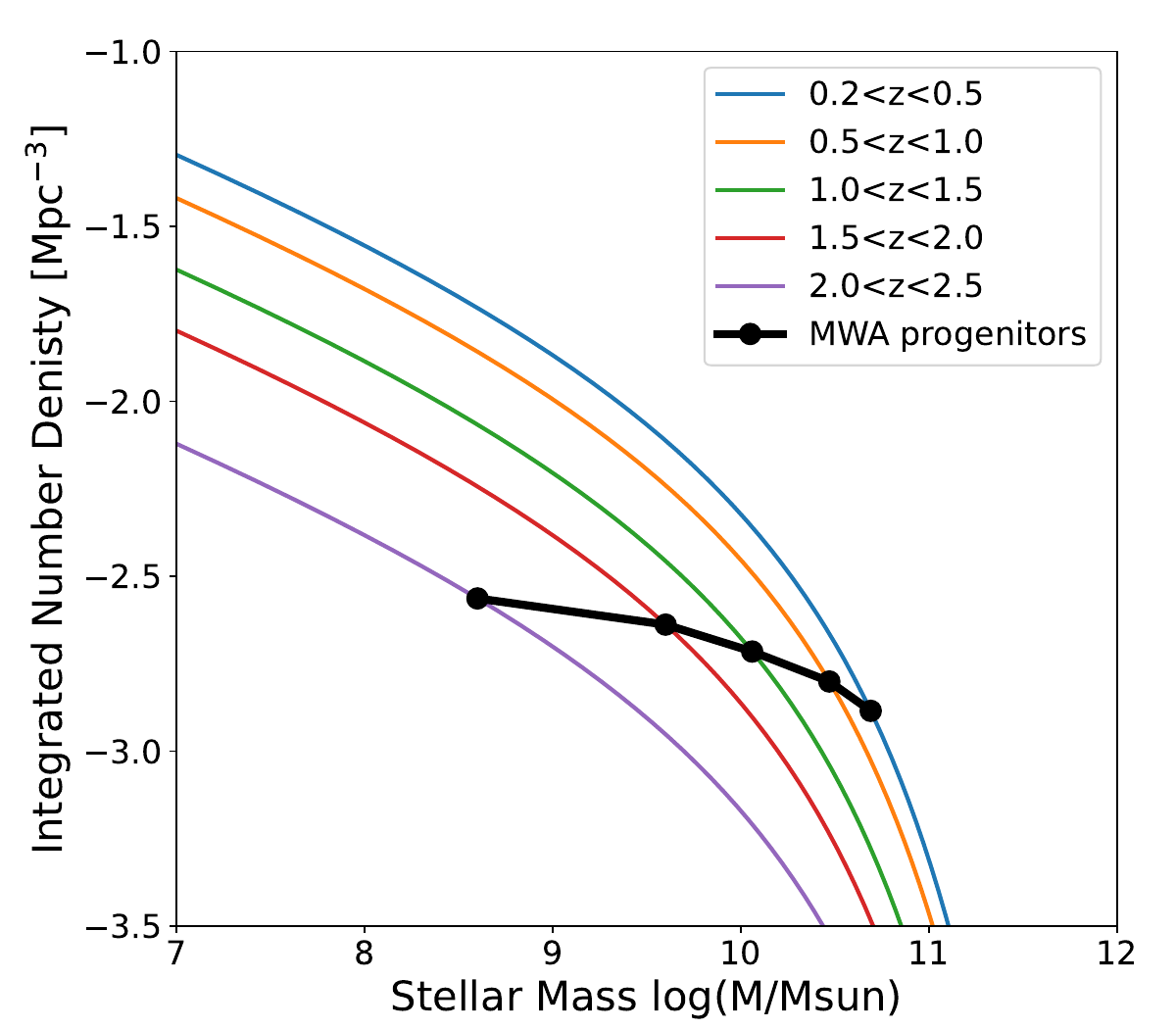}
    \caption{Integrated number density of galaxies from stellar mass functions (SMFs) from \cite{Muzzin:2013b}, and each black point is the median stellar mass of MWAs at the redshift range represented by each SMF.  The number densities in black are calculated using the \cite{Behroozi:2013b} code and are also shown in the right panel of Figure \ref{fig:num_density}. }
    \label{fig:schechter}
\end{figure}

In order to perform abundance matching to select progenitors of MWAs based on stellar mass, the stellar mass functions (SMFs) are needed. The stellar mass functions used in this paper come from the COSMOS/UltraVISTA Survey \citep{Muzzin:2013a}. The SMFs were measured from a sample of over 95,000 $K_s$-selected galaxies up to $z = 4$. They used 30 photometric bands to cover a wavelength range of $0.15\micron - 24\micron$. In addition to photometry from UltraVISTA \citep{McCracken:2012}, the catalog also includes additional data from GALEX, Subaru/SuprimeCAM, CFHT, and Spitzer. 

\section{Methods} \label{sec:methods}

\subsection{Abundance matching}
Our goal is to connect progenitors of MWA over cosmic time to measure their stellar mass growth. To do this, we adopt the semi-empirical approach using abundance matching presented in \cite{Behroozi:2013a}. The Behroozi code takes into account merger rates of galaxies at higher redshifts such that the comoving number density shifts to a slightly higher number density as redshift increases. These merger trees come from simulations of dark matter halos that the galaxies inhabit. Their code matches the rank order of galaxies according to stellar mass or luminosity with dark matter halos in decreasing order of peak historical halo mass. For our stellar mass range, the comoving number density increases $\lesssim 0.1$ dex for every 0.5 increase in redshift, in comparison to assuming a constant number density throughout cosmic time. 

We begin by determining the initial comoving number density for galaxies assuming the stellar mass of the Milky Way to be $10^{10.77}\msun$. This stellar mass value is generally within the accepted range of $ 10.5 \lesssim \log10(\mstar/\msun)\lesssim 11$, albeit on the higher end \citep{deRossi:2009,Bland-Hawthorn:2016,Sotillo-Ramos:2022}. This mass agrees with values for the stellar mass and initial comoving number density given in \cite{Marchesini:2009} for a MW-mass galaxy, which VD13 used in order to determine their MWA progenitors. This mass also corresponds with a comoving number density of $\log(\rho) = -2.95 \log(\text{Mpc}^{-3})$, which is similar to the density used in P15 for their MW progenitor sample.

Figure \ref{fig:num_density} plots the range of number densities and halo masses at higher redshifts for MWAs, with the solid lines indicating the median values, and the shaded blue regions indicating the 1-$\sigma$ deviation. This figure demonstrates how the code from \cite{Behroozi:2013a} determines the possible previous number densities at higher redshifts, given an initial number density at $z=0$. It is also able to determine the possible previous halo masses from the initial number density, and vice versa, using a relation between the cumulative number density to halo mass. The range of possible number densities and halo masses increases with redshift because the code from \cite{Behroozi:2013a} takes into account merger trees of halos. This means that a halo that exists at a given current number density at $z=0$ could have formed either from one particular halo in the past growing in isolation, or from mergers of smaller halos. 

In order to find the stellar masses of the MWA progenitors from the given previous number densities, they must be matched up to stellar mass functions from large scale observations. Figure \ref{fig:schechter} shows the method of obtaining the median stellar mass of the MWA progenitors using the stellar mass functions (SMFs) from the UltraVISTA survey from \cite{Muzzin:2013b}. For each of the SMFs that correspond to a range of redshifts, a stellar mass is matched to each of the median number densities predicted for that redshift by the code from \cite{Behroozi:2013a}. The minimum and maximum stellar mass for each redshift bin is taken to be the median redshift $\pm 0.2$ dex, which gives a more conservative estimate for the stellar mass range than using the corresponding stellar masses obtained from the SMFs with the number densities $\pm 1\sigma$.  Performing the same exercise with mass functions from \cite{Davidzon:2017} results in the progenitor masses being different by $0.01 - 0.14$ dex, making the difference in dex $<0.2$ dex at all redshifts.

Our total sample of galaxies is 110 within the redshift range of $0.3 < z < 2$, after the removal of  16 galaxies from the cluster fields for having magnification values exceeding 2.5 as computed from the \cite{Bradac:2009} lensing models, and a further removal of 18 galaxies from both cluster and flanking fields combined, from inspection of faults in their stellar mass maps (i.e. segmentation map falsely separated one object into two, too low signal to noise for enough spatial bins). There are 95 galaxies in the parallel fields and 15 galaxies in the cluster fields.

\begin{figure*} 
	\begin{subfigure}
        \centering
		\includegraphics[width=\textwidth]{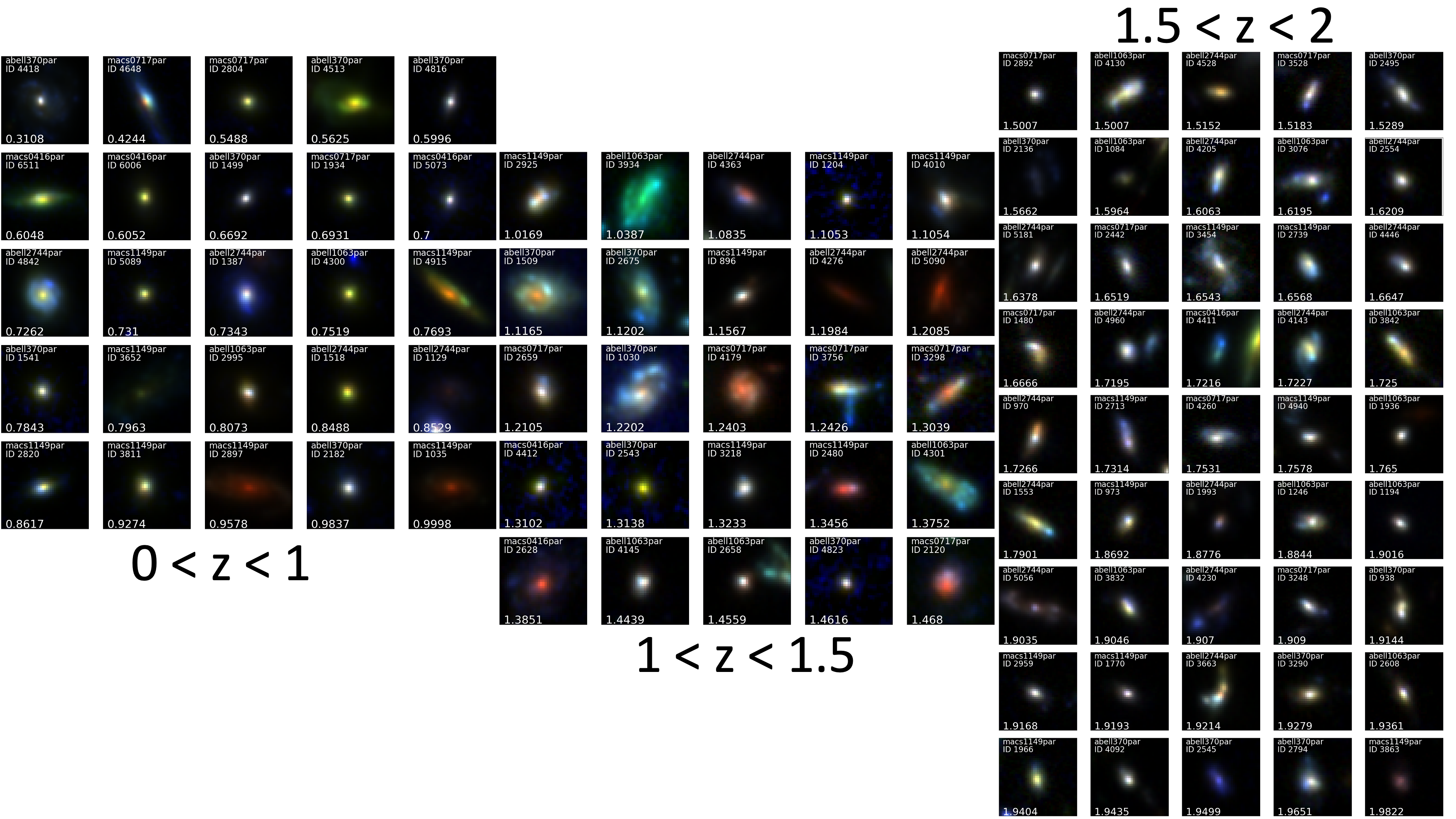}
	\end{subfigure}
	\begin{subfigure}
        \centering
		\includegraphics[width=\textwidth]{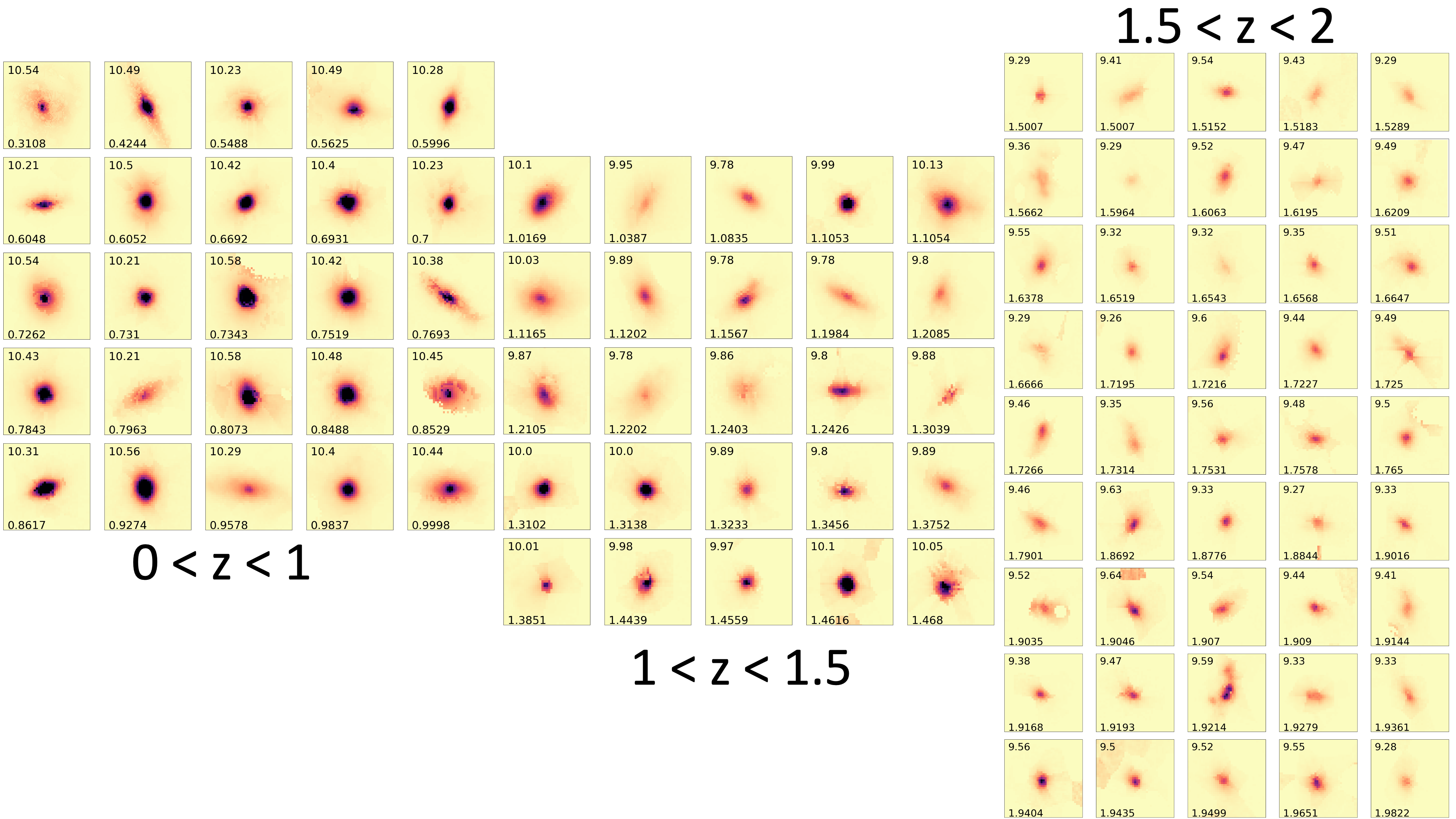}
	\end{subfigure}
	\caption{\emph{Top:} Colour images of MW progenitors in the HFF DeepSpace parallel flanking fields at different redshifts.
            \emph{Bottom:} Resolved stellar mass maps of the same MW progenitors. The min-max scale for the mass maps has the maximum set to $2\times10^8\msun$. The redshift of the sources is displayed on the bottom left of each thumbnail.}
 \label{fig:all-galaxies}
 \end{figure*}

 \begin{figure*}
	\begin{subfigure}
        \centering
		\includegraphics[width=0.49\textwidth]{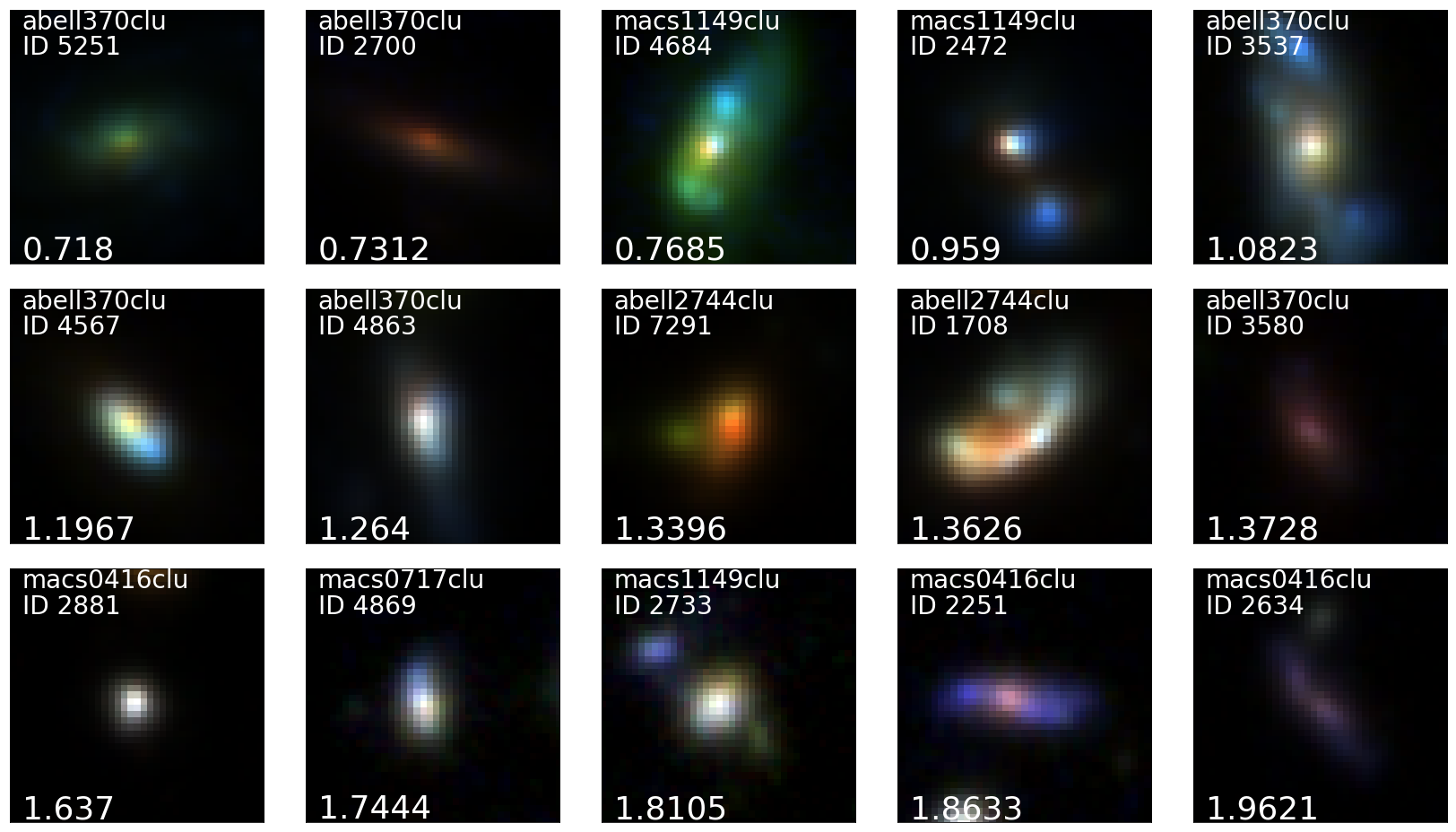}
	\end{subfigure}
	\begin{subfigure}
        \centering
		\includegraphics[width=0.49\textwidth]{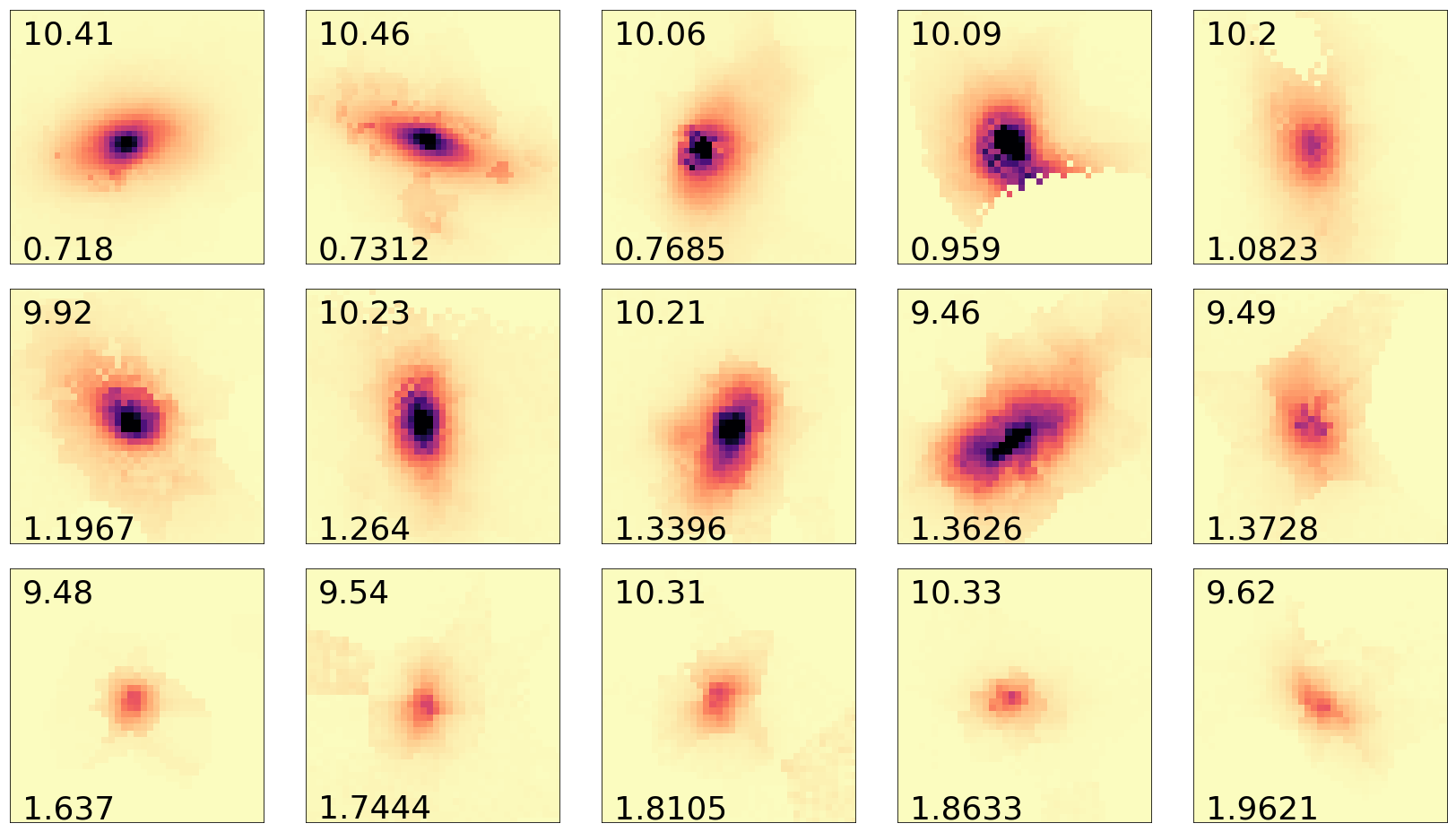}
	\end{subfigure}
	\caption{\emph{Left:} Colour images of MW progenitors in the HFF DeepSpace cluster fields at different redshifts. Every galaxy is beyond the cluster's redshift, and also has magnification effects of $< 2.5$.
            \emph{Right:} Resolved stellar mass maps of the same MW progenitors. The min-max scale for the mass maps has the maximum set to $2\times10^8\msun$. The redshift of the sources is displayed on the bottom left of each thumbnail.}
 \label{fig:all-galaxies2}
 \end{figure*}

 \begin{figure*}
 \begin{subfigure}
        \centering
		\includegraphics[width=0.49\textwidth]{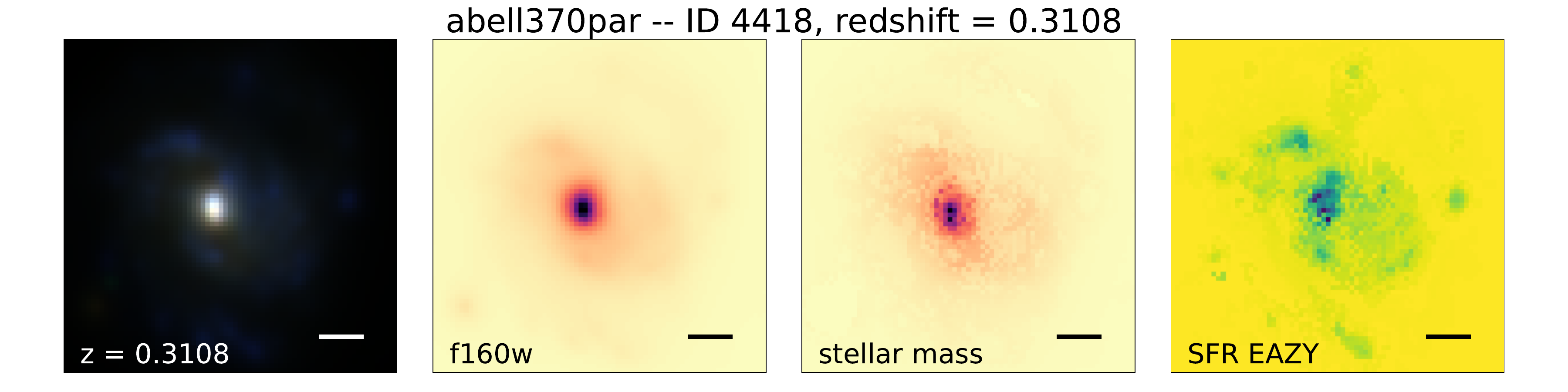}
	\end{subfigure}
 \begin{subfigure}
        \centering
		\includegraphics[width=0.49\textwidth]{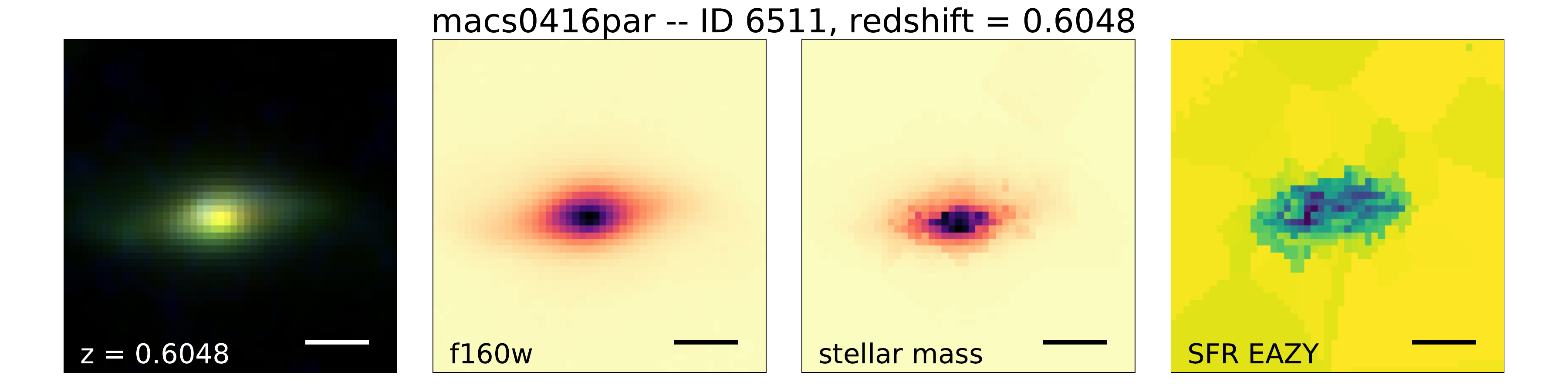}
	\end{subfigure}
 \begin{subfigure}
        \centering
		\includegraphics[width=0.49\textwidth]{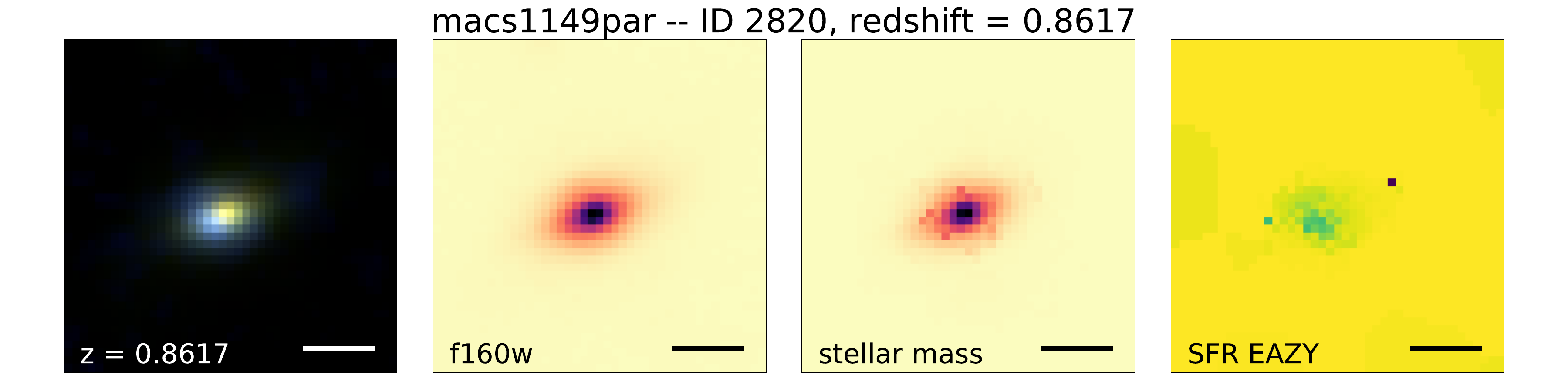}
	\end{subfigure}
 \begin{subfigure}
        \centering
		\includegraphics[width=0.49\textwidth]{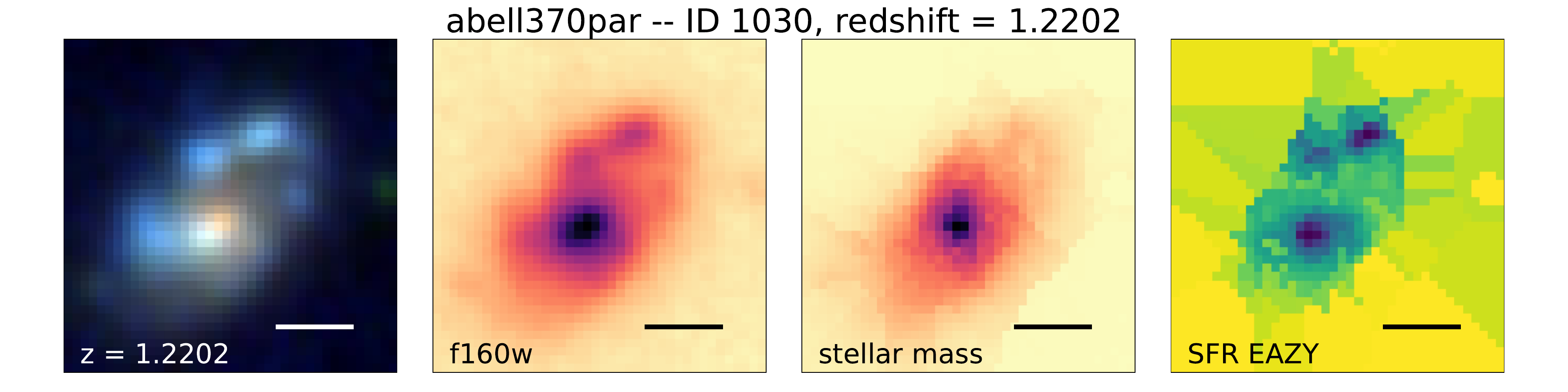}
	\end{subfigure}
 \begin{subfigure}
        \centering
		\includegraphics[width=0.49\textwidth]{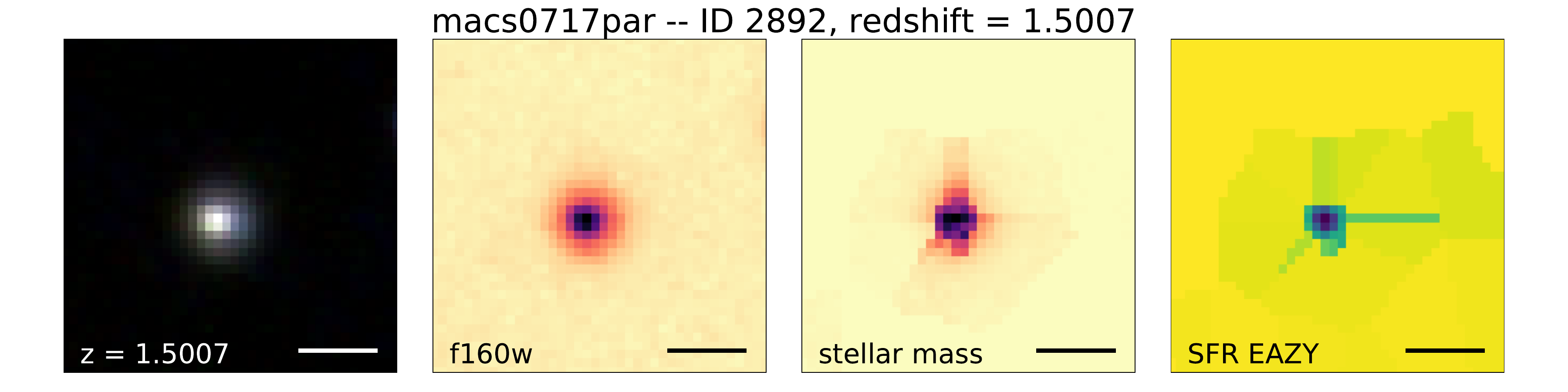}
	\end{subfigure}
 \begin{subfigure}
        \centering
		\includegraphics[width=0.49\textwidth]{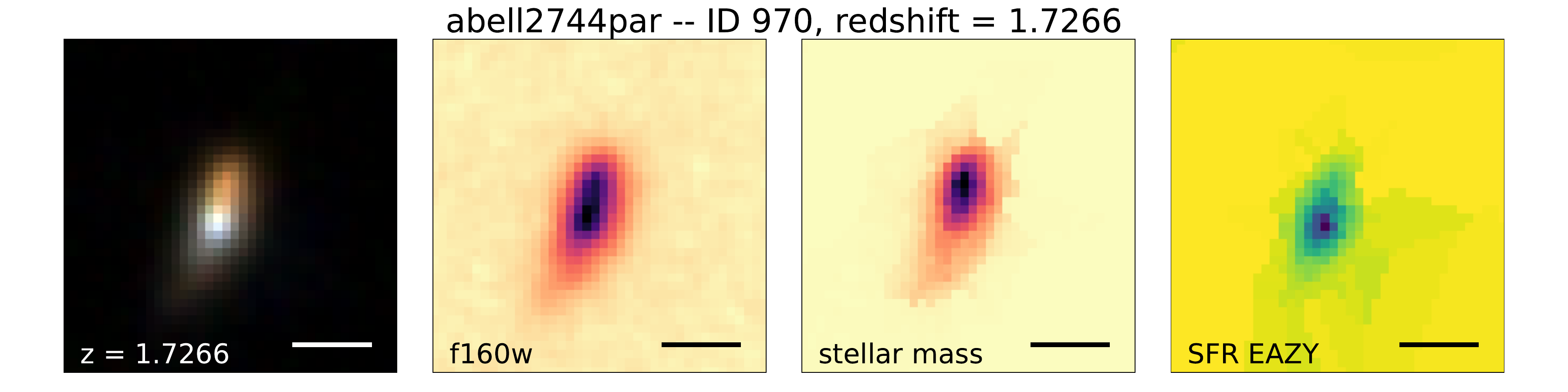}
	\end{subfigure}
    \begin{subfigure}
        \centering
		\includegraphics[width=0.49\textwidth]{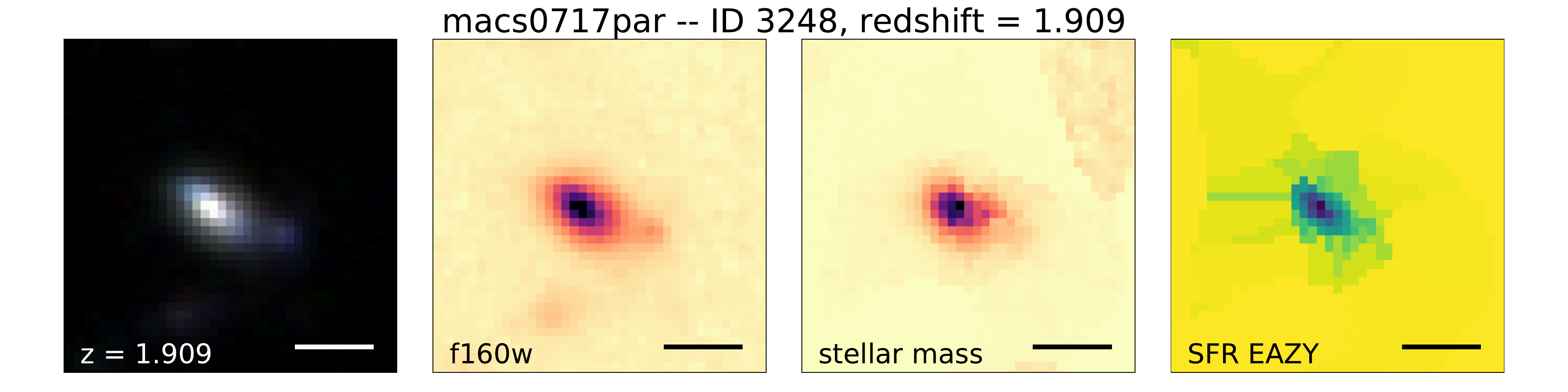}
	\end{subfigure}
 \begin{subfigure}
        \centering
		\includegraphics[width=0.49\textwidth]{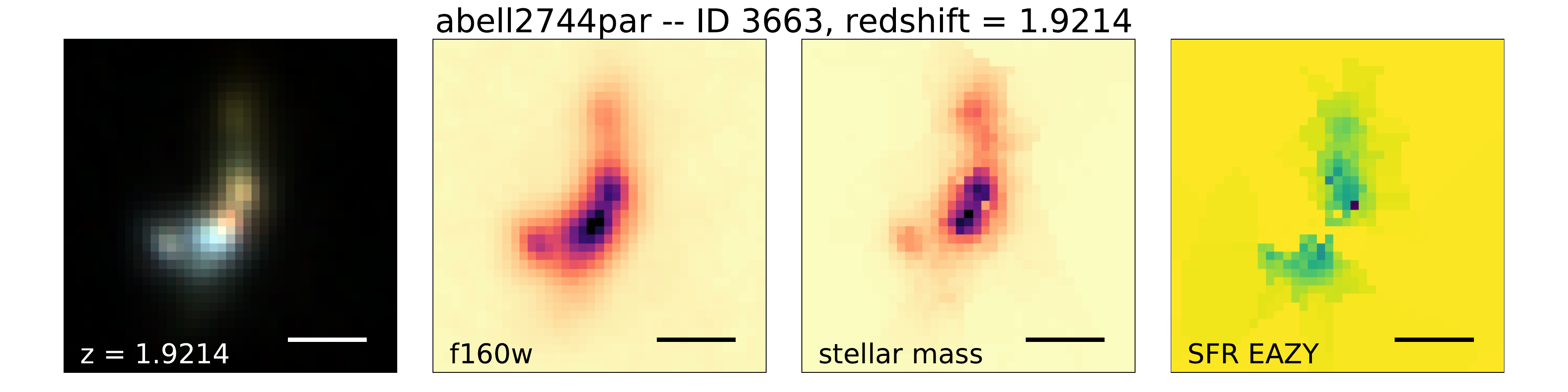}
	\end{subfigure}
 
	\caption{Comparison of colour image, F160W , stellar mass, and SFR profiles. The horizontal line in the bottom right represent $3\times$FWHM of the F160W PSF. The the third object from the top, in the left column(MACS0717PAR ID 2892) is the most compact object in terms of its effective radius respective to the PSF's FWHM at that redshift.}
 \label{fig:galaxy-comparisons}
 \end{figure*}

\begin{figure}
    \centering
    \includegraphics[width=\columnwidth]{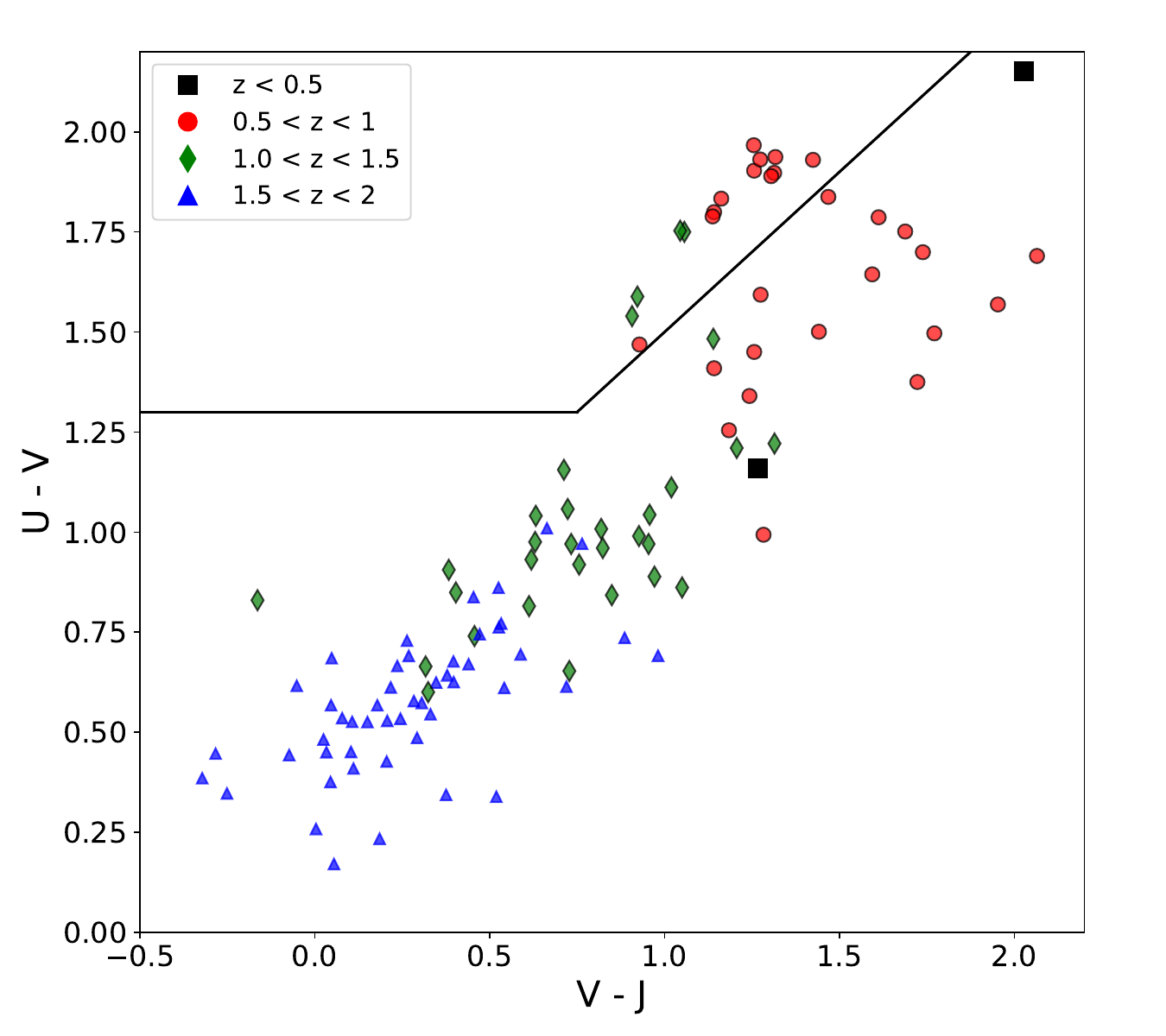}
    \caption{UVJ Diagram with boundaries from \cite{Shipley:2018}. There are 11 quiescent galaxies at redshift $0.5 < z < 1$, and 4 quiescent galaxies at redshift $1 < z <1.5$.}
    \label{fig:uvj}
\end{figure}

\begin{figure*}
    \centering
    \includegraphics[width=\textwidth]{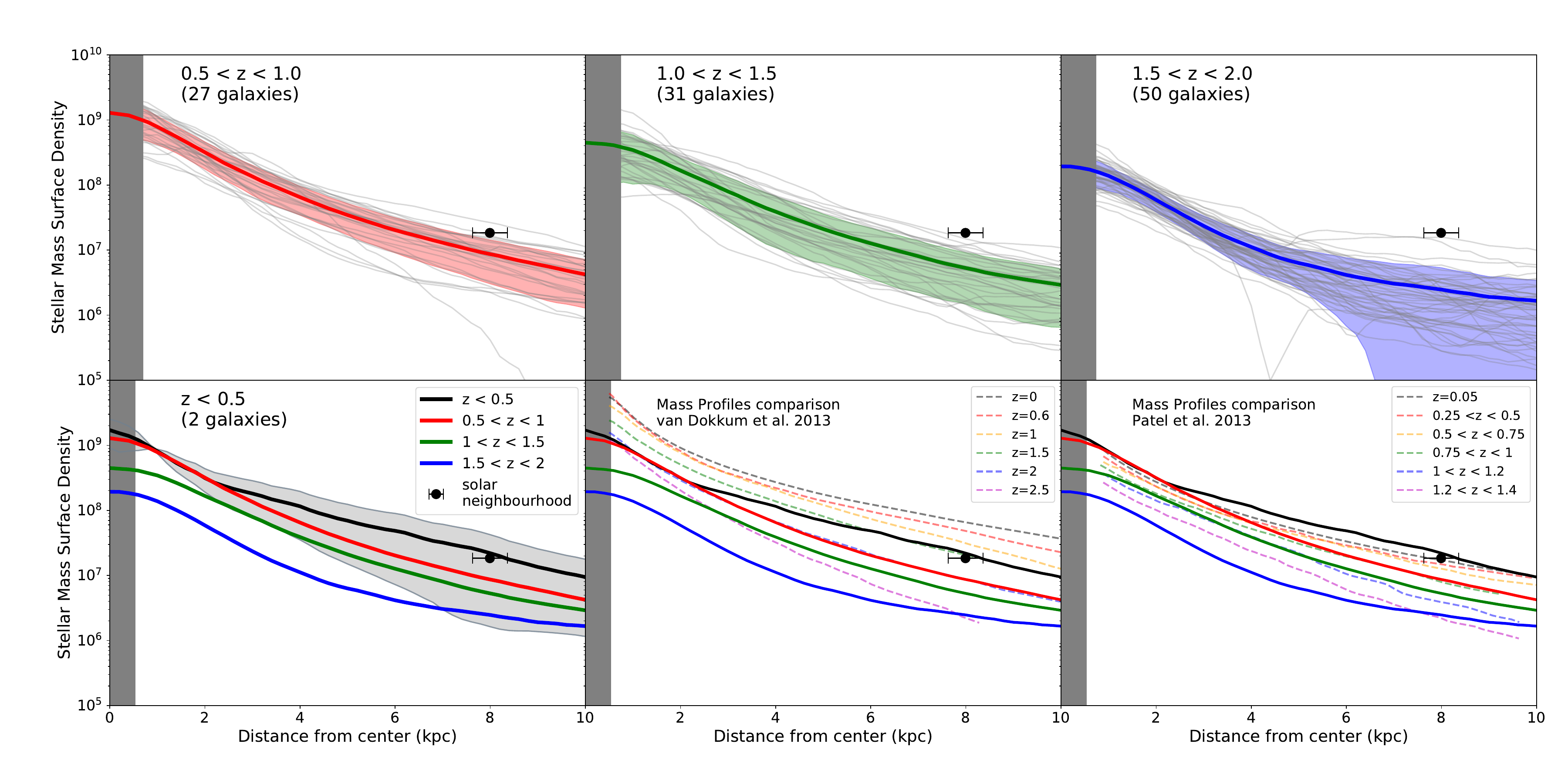}
    \caption{\emph{Top Row and Bottom Left panel:} Every surface mass density profile created from stacking and normalizing each 1-D stellar mass profile at each redshift bin. Light grey lines represent the individual stellar mass profiles in each redshift bin. Shaded regions in the same colour of the profile represent the 1-$\sigma$ deviation from the stacked profile. The bottom left panel also contains normalized 1-D surface mass profile for \textit{each} redshift. The grey lines are the two galaxies which are in the redshift bin $0 < z < 0.5$, with the shaded grey region representing the 1-$\sigma$ deviation. The black point is the local solar neighbourhood stellar mass density from \cite{Bovy:2012}. The profiles are integrated up to $r = 20$kpc. 
\emph{The bottom center and right panels:} These panels compare our derived profiles at each redshift bin with previous results. Our mass profiles agree well with the profiles from P13 more than VD13, despite using abundance matching like the latter, instead of the star-formation main sequence like the former. The vertical grey regions in every panel indicates the part of the profile affected by the PSF.}
    \label{fig:sm-profiles}
\end{figure*}

\subsection{Stellar mass and SFR map construction}

In Figures \ref{fig:all-galaxies} and \ref{fig:all-galaxies2}, we display the colour images and the stellar mass maps of all 110 galaxies in our sample. Figure \ref{fig:all-galaxies} contains all 95 MWA progenitors from the parallel fields and Figure \ref{fig:all-galaxies2} contains the 15 progenitors from the cluster fields. The colour images are constructed from the filters that are the closest match to rest-frame $U$-band , rest-frame $V$-band, and the $F160W$ filter. 

Figure \ref{fig:galaxy-comparisons} shows the colour image, F160W cutout, stellar mass map, and SFR map of several galaxies in our sample of various compactness. As stated before, the images are all PSF-matched and convolved to the WFC3/F160W resolution to ensure consistent resolution in each filter. The angular resolution is $0.25 - 0.50$ kpc per pixel and $0.8 - 1.5$. kpc per PSF FWHM, for the redshift range $0.3 < z < 2$. From the HFF DeepSpace catalogs, the smallest effective radius in our sample is 3.27 pixels, and the largest effective radius is 18.1 pixels. The size of the FWHM of the PSF in pixels is 2.95, so using 2 times the effective radius as the size of the galaxy in pixels, the most compact galaxy is 2.2 times the PSF FWHM, and the most extended galaxy is 12.3 times the PSF FWHM. The effective radii are based on the detection image, which is a combination of the reddest filters from F814W to F160W as measured by \texttt{SExtractor}.  

The first step to generating resolved stellar mass profiles as well as resolved SFR profiles is to spatially bin the images containing the photometry of the MWA progenitors. We apply the Voronoi tessellation algorithm from \cite{Cappellari:2003}, and SED-fitting with FAST++ \citep{Kriek:2009, Schreiber:2015}, in order to generate the stellar mass maps of the MWA progenitors from their photometry. The reason why we chose the Voronoi tessellation method for pixel-binning is due to the geometry of the tessellated bins being able to best homogenize the SNR throughout the entire image of the galaxy, and in a way which preserves maximum spatial resolution in the geometry of the shape of the bins. For each galaxy, a lower limit for SNR of 10 is placed on each bin. But most galaxies at $z < 1.5$ have enough signal to noise that an SNR lower limit of 30 is used instead. 

In the SED-fitting of the photometry of the selected galaxies, we use BC03 stellar populations \citep{Bruzual:2003}, a Chabrier IMF \cite{Chabrier:2003}, a Calzetti dust law \citep{Calzetti:2000}, and a delayed tau model SFH. The SED-fitting code FAST is used due to the accuracy and robustness of the stellar mass outputs. We note that using a Calzetti dust law and a delayed tau model SFH may not be accurate to the star-formation history of the Milky Way; delayed tau models allow only for a rising SFH at the very beginning, but have a declining SFH for the vast majority of cosmic time. In the redshift range probed by this work however, a declining SFH is a reasonable assumption, since the global trend of cosmic SFR has been declining since $z\sim2$.

Once a stellar mass is obtained for each spatial bin, we construct the stellar mass map for that object. The maps have the degraded resolution of the Voronoi tessellation, however we regain the original resolution by an additional scaling applied to each pixel. We distribute the amount of stellar mass within one bin to each of its constituent pixels according to the pixel's contribution to the F160W flux. For more information on the details of resolved stellar mass map construction, we refer the reader to \cite{Tan:2022}.

In addition we have also created 2-D star-formation rate maps. The SFR profiles were \textit{not} derived from FAST SED-fitting, because while assuming a delayed tau model SFH results in reasonable stellar mass derivations up to $z \sim 3$ (see appendix of \citealt{Muzzin:2013b}), it \textit{does not} result in the most accurate SFRs for galaxies at higher redshifts.  Therefore, we instead derive SFR from the integrated UV luminosity calculated using EAZY \citep{Brammer:2008}, by placing a UV-filter (1400-1600\AA) on the rest-frame fitted spectra and obtaining the total rest-frame UV luminosity. To lessen the effect of parameters on the resulting fits, we still used a delayed tau SFH, a Chabrier IMF, and a Calzetti dust law. However, the template libraries were from FSPS \citep{Conroy:2010}. We then apply a dust correction using the Calzetti dust law curve with the $A_v$ derived from the HFF DeepSpace catalog (i.e. the $A_v$ outputted by FAST while fitting for the integrated stellar mass of each object.) 

\begin{figure*}
	
		\includegraphics[width=\textwidth]{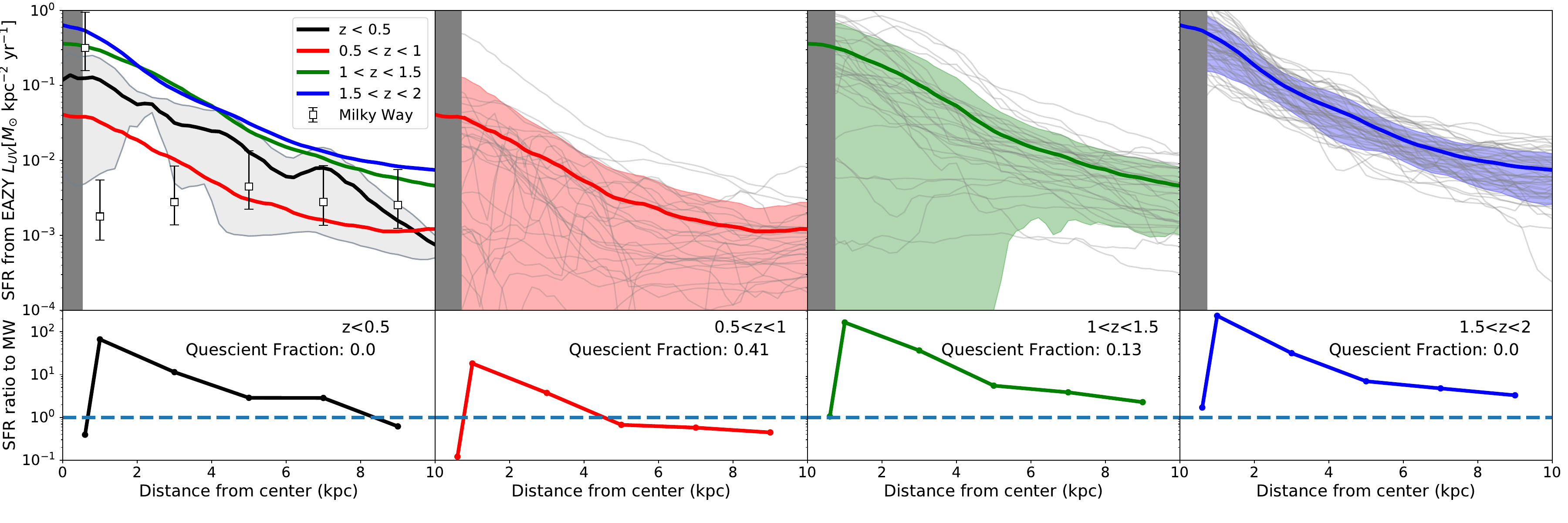}
	
\caption{\emph{Top Row:} Similar plots to \ref{fig:sm-profiles}, but for stacked and normalized SFR profiles. The same galaxies are in each redshift bin. The leftmost plot also has every stacked-normalized SFR profile from each redshift plotted on top for a comparison. In addition, the white points in the leftmost panel indicate the Milky Way's current day SFR density versus $R$, from \cite{Kennicutt:2012}.  The vertical grey regions in every panel indicates the part of the profile affected by the PSF.
 \emph{Bottom Row:} Ratio between the current Milky Way $\Sigma_{SFR}$ from \cite{Kennicutt:2012}, and the $\Sigma_{SFR}$ at the same $R$ for each past redshift bin. The dotted line indicates where the ratio is the same. Although magnitude of change in $\Sigma_{SFR}$ varies between redshifts, the overall shape remains consistent. }
 \label{fig:sfr_profiles}
 \end{figure*}

Once the stellar mass maps have been constructed, we obtain 1D stellar mass density profiles for each galaxy by placing down elliptical apertures/annuli centered on the galaxy of increasing radial distance (each are 0.2 kpc in width) up to 20 kpc distance in the major axis, and then taking their stellar mass surface density. The same process is also used on the resolved SFR maps to create 1D SFR profiles, also up to 20 kpc. The major and minor axes, as well as the position angles used to determine the placement of the annuli are from the HFF DeepSpace photometric catalog, extracted from their detection filters which utilized a combination of $F814W$, $F105W$, $F125W$, $F140W$, and $F160W$ bands.

This makes it a fair comparison with the light profiles in the works of P13 and VD13 because the former also integrated the profiles out to 20 kpc and the latter out to 25 kpc. We also mask out nearby galaxies from the 2D profiles before creating the 1D profile so their mass or SFR does not interfere with the surface density calculations. We use a UVJ diagram (see Figure \ref{fig:uvj}) to examine which galaxies are star-forming or quiescent in our sample. The UVJ boundaries are from \cite{Shipley:2018}. Out of the 110 total galaxies from all twelve cluster and parallel fields, only 15 galaxies lie within the quiescent region of the UVJ diagram. A small number of  $z < 1$ galaxies are in the upper-right region of the UVJ diagram, indicating the presence of dust, although this can be an effect of viewing angle. The majority of galaxies, especially at $z > 1$ (green and blue points), are not dusty and highly star-forming.

For each redshift, we stacked and renormalized the 1-D mass profiles based on their total stellar masses from the profile, to create the averaged 1-D mass profile at that redshift range. The same was done for the star-formation rate maps in order to obtain the normalized 1-D SFR profiles at each redshift bin. The normalized mass profiles are plotted in Figure \ref{fig:sm-profiles}, and the normalized SFR profiles are plotted in Figure \ref{fig:sfr_profiles}. 


We expect the MWA progenitors to be highly star-forming as redshift increases, and that is reflected in the sample. The Milky Way is currently in the green valley, either because of its dust content or because it is in the process of transitioning from star-forming to quiescent.

We note that the quiescent fraction is $41\%$ at redshift $0.5 < z < 1$, but only $13\%$ at redshift $1 <z < 1.5$, and all MW progenitors are star-forming at $1.5 < z < 2$. The redshift bin $z < 0.5$ only contains two galaxies, and both are star-forming according to the UVJ diagram in Figure \ref{fig:uvj}. Therefore, the galaxies being majority star-forming disk galaxies at higher redshifts matches the canonical picture of the evolutionary history of the Milky Way. \citep{Bland-Hawthorn:2016}.  Given that the Milky Way is currently in the green valley with a modest SFR, that $\sim$ 60\% of its progenitors (by stellar mass) at 0.5 $< z <$ 1.0 are star-forming and $\sim$ 40\% are quiescent is also reasonable.

For our sample, we used GALFIT \citep{Peng:2002} to fit the stellar mass maps in order to obtain the S\'ersic indices and half-mass radii in the same procedure as outlined in \cite{Tan:2022}. We fit for S\'ersic index, effective radius, axis ratio, and position angle, though the parameters of interest are only the first two. Initial guesses for the parameters were \texttt{SExtractor} values \citep{Bertin:1996} for effective radius, semiminor/semimajor axis, and position angle for each object from the HFF DeepSpace catalog, and $n=2$ for the S\'ersic index, but with constraints of $n=0$ to $n=8$. We will compare their redshift evolution with VD13, P13, and P15 in \S\ref{sec:results-galfit} 

\section{Results} \label{sec:results}
\begin{figure*}
    \begin{subfigure}
        \centering
		\includegraphics[width=\textwidth]{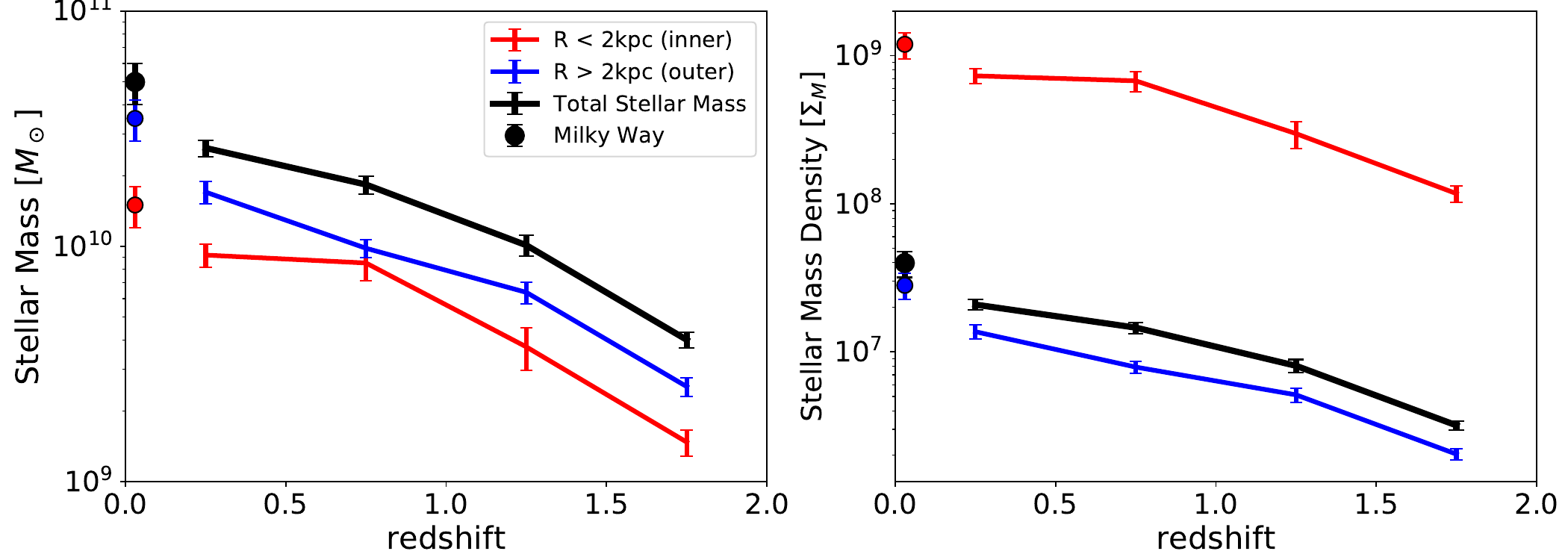}
    \end{subfigure}
    \begin{subfigure}
        \centering
        \includegraphics[width=\textwidth]{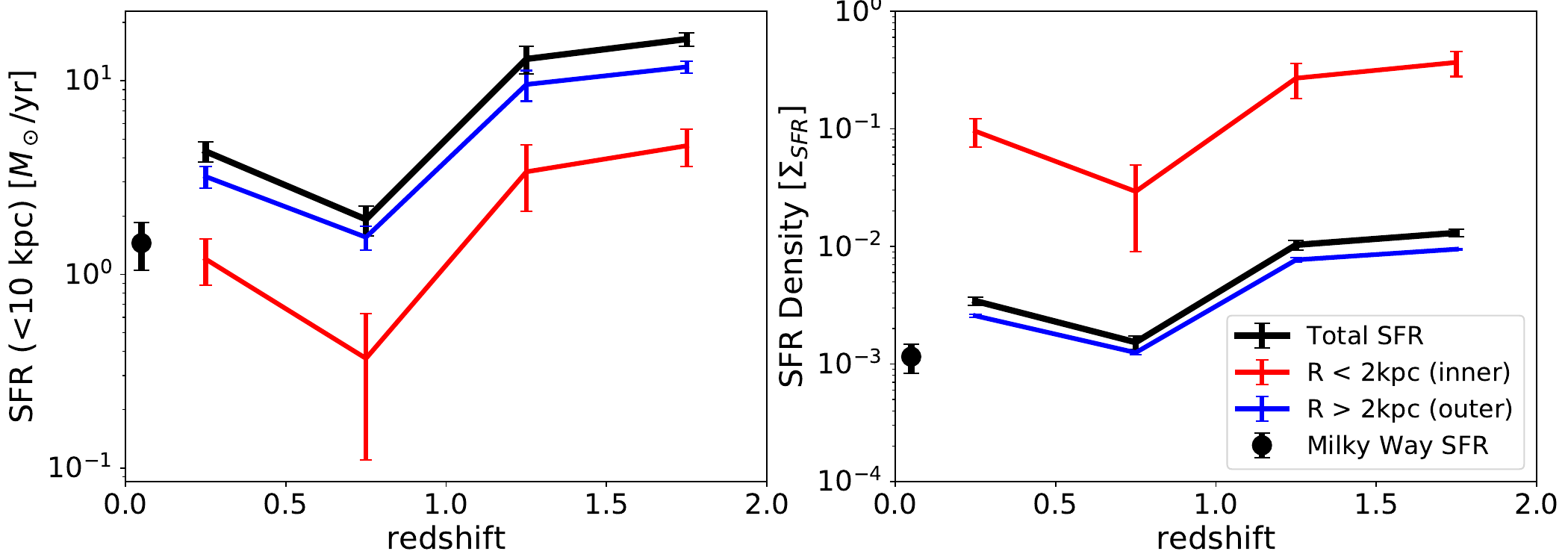}
    \end{subfigure}
    \caption{\emph{Top row:} Evolution of the total stellar masses (left) and stellar mass densities ($\Sigma_{\star}$, right) of MWA progenitors out to $z\sim 2$. $\Sigma_{\star}$ for each region is calculated by taking the stellar mass and dividing it by the surface area of its respective regions. Black lines represent the galaxy-integrated properties out to 20 kpc, whereas red and blue lines show the inner ($R<2$ kpc) and outer regions ($R > 2$ kpc), respectively. The Milky Way $\Sigma_\star$ were taken from the measurements of the bulge stellar mass, total stellar mass, and $B/T$ ratio from \cite{Bland-Hawthorn:2016}, but divided by the surface area to give a surface mass density. This is assuming the bulge is contained within a galactocentric radius of 2 kpc, and that the disk integrates out to 20 kpc, for consistency with MWA progenitor mass profiles.
\emph{Bottom row:} Similar to plots above, but showing the evolution of total SFRs (left) and SFR densities ($\Sigma_{SFR}$, right). The Milky Way $\Sigma$SFR is again derived from the total SFR in \cite{Bland-Hawthorn:2016} but divided by surface area. }
   \label{fig:mass-assembly}
\end{figure*}
\subsection{1-D Stellar Mass Profiles}

 In Figure \ref{fig:sm-profiles}, the stacked and normalized stellar mass surface density profiles for each redshift bin are plotted as a function of galactocentric radius $R$. Our main result is that these stacked and normalized $\Sigma_\star$ profiles have approximately the same overall shape up to 10 kpc. In the bottom left panel where the  $\Sigma_\star$ profiles for each epoch are plotted together, it is clear that the growth in $\Sigma_{\star}$ between different epochs do not depend on $R$ to a great extent.  With the stellar mass density of the solar neighbourhood (from \citealt{Bovy:2012}) overlaid on top, it is evident that by redshift 0.5, most of the stellar mass has already been built up for this sample of MWA progenitors. Even before $z = 0$, these galaxies have been able to reach the stellar mass density of the Milky Way. Although we note with only two galaxies in the redshift bin of $0 < z < 0.5$, this result could benefit from a larger sample.

At each redshift step from $z \sim 2$ to $z\sim 0$, the inner 2 kpc region appears to grow more at earlier times.  The stacked $\Sigma_\star$ profile does not change at $R \lesssim 3$kpc  between $z < 0.5$ (the black curve) and $0.5 < z < 1$ (the red curve). The profiles at around 8 kpc, which is the distance the solar neighbourhood is from the MW's galactic centre, increase in mass density more at earlier times ($z > 1$), and less so at later times ($z < 1$). It increased from a mass density of $\log(\mstar/\msun)\text{kpc}^{-2}=6.42$ to $\log(\mstar/\msun)\text{kpc}^{-2} = 6.81$, which is a difference of 0.35 dex, but from $z\sim 1$ to $z\sim 0.5$, it only increased in mass density to $6.94 \log(\mstar/\msun)\text{kpc}^{-2}$, a difference of 0.13 dex. To characterize the mass build-up with more accuracy, \S\ref{sec:inner-outer-growth} charts the change in total stellar mass and stellar mass density $\Sigma_\star$ in more detail.

For comparison, we have plotted our 1-D stellar mass profiles against the mass profiles of previous works, which is displayed in the bottom center and bottom right panels of Figure \ref{fig:sm-profiles} against VD13 and P13 respectively. Our stellar mass profiles are on a similar order of magnitude to P13's results, whereas VD13's stellar mass profiles are at least 1 dex greater than ours across all radii. This is slightly counter-intuitive, because both VD13 and P15 use a starting mass for the Milky Way of $5\times10^{10} \msun$ or $10^{10.7} \msun$, and P13's starting mass is $10^{10.5} \msun$, which is around $3\times10^{10} \msun$. Our starting mass of $10^{10.77} \msun$ is actually closer to the former, but we find similar MWA progenitor masses to P13. Our mass profile at $0.5 < z < 1$ follows closely P13's density profiles at $0.7 < z< 0.1$ from $R\sim 4$ kpc to $R\sim10$ kpc. At $R \lesssim 4$ kpc, the lowest redshift mass profile shows that the stellar mass density is slightly ($\sim 0.1$ to $0.3$ dex) above P13's at a similar redshift range.
Our mass profile at $1 < z < 1.5$ falls between P13's profiles at similar mass ranges, up until $R \gtrsim 8$ kpc, when our profile flattens out above P13's $1 < z < 1.2$ profile. In general, the shape of our mass profiles are less steep in the centre and level off at a higher mass density.

\subsection{1-D star formation rate profiles}

The other important factor to stellar mass growth is the star-formation rate. We make 1-D SFR profiles, representing the SFR versus galactocentric radius in kpc, and these are shown in Figure \ref{fig:sfr_profiles} on the top row. In the top-left panel of Figure  \ref{fig:sfr_profiles}, the star formation rate density profiles show more star formation within the $R < 2 $ kpc region, but there are proportionally more stars there. Overall the distribution of the SFRs and stellar mass shows that the increase in stellar mass due to star formation in each redshift range would be approximately proportional to the stellar mass that is already present. 
\newline\indent
Figure \ref{fig:sfr_profiles} shows that the SFRs are overall more than 1.3 dex higher from $ 1 < z < 2$ as compared to $z < 1$. 
The sharp drop in overall SFR between these redshifts might be expected as the SFRs of all galaxies start to decline strongly at $z < 1$.  Interestingly for the MWA progenitors, the drop in SFR is uniform over the entire profile, implying that the mass growth from star formation is not preferentially in the inner or outer regions of MWAs. This in conjunction with the stellar mass profiles indicate that mass assembly happens in lockstep from $z \sim 2$ onwards, and also slows down in lockstep as well. 

In addition to the stacked and normalized SFR profiles for each redshift bin, in the bottom row of Figure  \ref{fig:sfr_profiles} we also plot the ratio between the Milky Way's measured SFR density against the SFR density at the same $R$ for each of the stack and normalized SFR density profiles. Although the magnitude of the ratio of $\Sigma_{SFR}$ changes, the overall shape of this ratio of past $\Sigma_{SFR}$ to current day MW $\Sigma_{SFR}$ does not change qualitatively. This has implications for bulge-growth, as this implies a decrease in $\Sigma_{SFR}$ over time occurred at similar rates across the entire galaxy, with no preference for inner or outer regions. Therefore it is another example of lockstep growth in both the bulge and the disk regions of MWA progenitors.

\subsection{Growth in inner versus outer regions}\label{sec:inner-outer-growth}
\begin{figure}
    \includegraphics[width=\columnwidth]{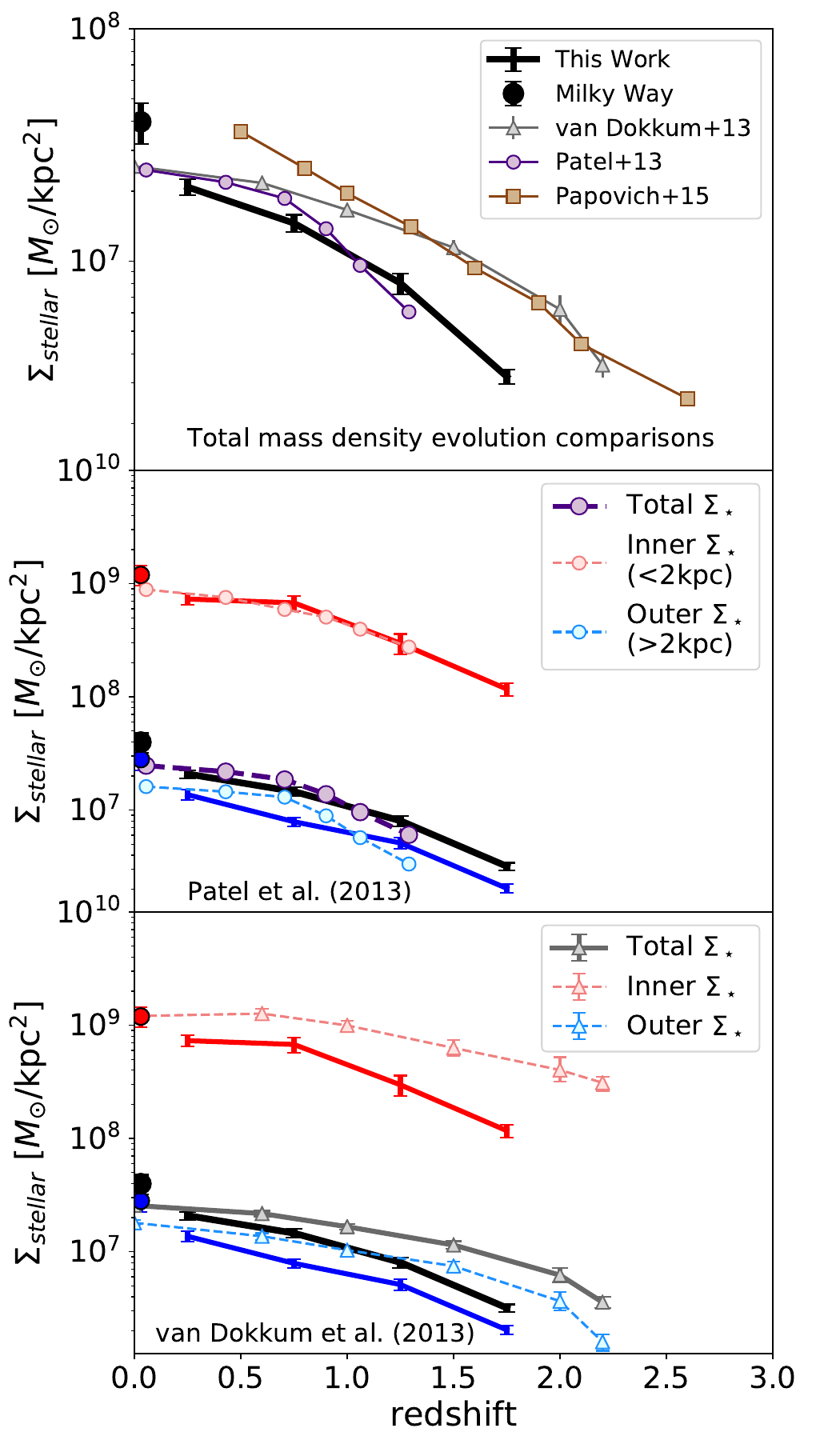}
\caption{\emph{Top Panel:} Comparison of the \textit{total} $\Sigma_\star$ as a function of redshift  between our densities and P13, VD13, and P15. \emph{Middle panel:} The inner and outer stellar mass densities of our results and P13. \emph{Bottom panel:} Same regional stellar mass density comparisons as the middle panel but with VD13's densities. MW mass density points are again derived from measurements listed in \cite{Bland-Hawthorn:2016} as in Figure \ref{fig:mass-assembly}.}\label{fig:sm-growth-compare}
\end{figure}
\subsubsection{Mass assembly in the inner and the outer regions}

To further investigate the stellar mass assembly in each redshift range, we divide the growth into inner vs.~outer regions. The inner region is defined as within 2 kpc of the center of each galaxy's profile, and outer region is defined as beyond 2 kpc. The profiles are integrated out to 20 kpc, to remain consistent with the limit used in P13. 

In Figure \ref{fig:mass-assembly}, we plot the stellar mass surface density $\Sigma_{\star}$ versus redshift in the top panels, and SFR density $\Sigma_{SFR}$ versus redshift for the bottom panels. These are obtained by integrating over the 1-D density profiles by $R$ up to 20 kpc.  In addition, we also plot the inner ($R < 2$kpc) and outer (2kpc $< R <$ 20kpc) $\Sigma_{\star}$ and $\Sigma_{SFR}$ against redshift, which are the red and blue lines respectively. This is not only to demonstrate overall mass assembly and change in SFR over time, but also whether the change was more pronounced towards the central regions or the outskirts of the galaxies for each epoch. 

Figure \ref{fig:mass-assembly} shows that the stellar masses of the inner regions ($R < 2$ kpc) are consistently around 0.5 dex lower than the outer regions at every redshift bin. Compared to the total mass and total mass density of the Milky Way from local measurements, the Galaxy is not a perfect extrapolation of the trends shown. In the top left panel of Figure \ref{fig:mass-assembly}, the outer regions of the Milky Way (the blue point at $z=0$) have more stellar mass compared to and extrapolation of the trend for the MWA progenitors' outer regions ($R > 2$ kpc) at higher redshifts (the blue line) to $z = 0$. Also the stellar mass of the inner 2 kpc of the Milky Way is similar to the total stellar mass of the MWA progenitors at $z \sim 0.5$. This seems to indicate the inner 2 kpc for the Milky Way stopped growing while the outer regions continued to grow, however, our result suggests this may not be the case for MWAs in general, if they follow the trends seen in the top two panels of Figure \ref{fig:mass-assembly}.

It is useful to note that that while our MWA progenitors do not indicate a similar type of growth from $z\sim0.5$ to $z\sim 0$ that would match the current Milky Way's properties, neither do the MWA progenitor samples of the previous works that they are compared with. If we took the full sample of all local galaxies considered \textit{``Milky Way Analogues"}, perhaps the Milky Way's higher than expected disk growth at low redshifts is anomalous among galaxy analogues of similar mass.

\subsubsection{Star-formation at different regions and different epochs}

In Figure \ref{fig:mass-assembly}, the bottom left panel plots overall median SFR versus redshift, and the bottom right panel plots SFR density versus redshift. Those panels demonstrate the overall trend of star-formation rate decreasing with decreasing redshift. However, the drop in median SFR is more drastic from $z \sim 1.5$ to $z < 1$ than any other epoch. This holds for both the inner and outer regions of the MWA progenitor galaxies, which can be seen in the SFR profiles in the bottom row of Figure \ref{fig:mass-assembly}. Since $z \sim 2$ is what is known as ``cosmic noon" when the peak of universal star-formation occurred, the decline may be expected. However it is interesting to note that the decrease in median SFR for the MWAs (for all regions) is less drastic in the redshift bin $1.5 < z < 2$ than the redshift bin $1 < z < 1.5$. 

The inner region ($R < 2$ kpc) and outer regions ($R > 2 $ kpc) of the the MWA progenitors have similar changes in SFR when moving from one redshift bin to another. This indicates lockstep star-formation rates for MWA progenitors since $z \sim 2$. This means that when the SFR changes, it changes at the same rate over \textit{the entire galaxy's surface area}, whether it is in the $R < 2$ kpc region or in the outer region at $R > 2 $ kpc. Generally, SFR is decreasing with decreasing redshift, but there is a small increase in the median SFR when moving from the redshift bin $0.5< z < 1$ to $0.3 < z < 0.5$, as seen in the bottom left panel of Figure \ref{fig:mass-assembly}. Looking instead to the bottom right panel of Figure \ref{fig:mass-assembly}, where the SFR density of the MWA progenitors is plotted versus redshift, there is more of an increase in the SFR density of the $R < 2$ kpc inner region than the outer region. This might be evidence of mild bulge-growth, but because it is minor, it does not break the general trend that SFR changes in lockstep across all radii.

\cite{Mosleh:2017} also finds evidence that star-forming galaxies have their central and outer regions grow concurrently. Their range of stellar masses were from  $10^{10.2}\msun$ to $10^{11.2}\msun$, so the Milky Way would be comparable to the star-forming galaxies in their highest mass bin. 

\begin{figure*}
    \centering
    \includegraphics[width=0.8\textwidth]{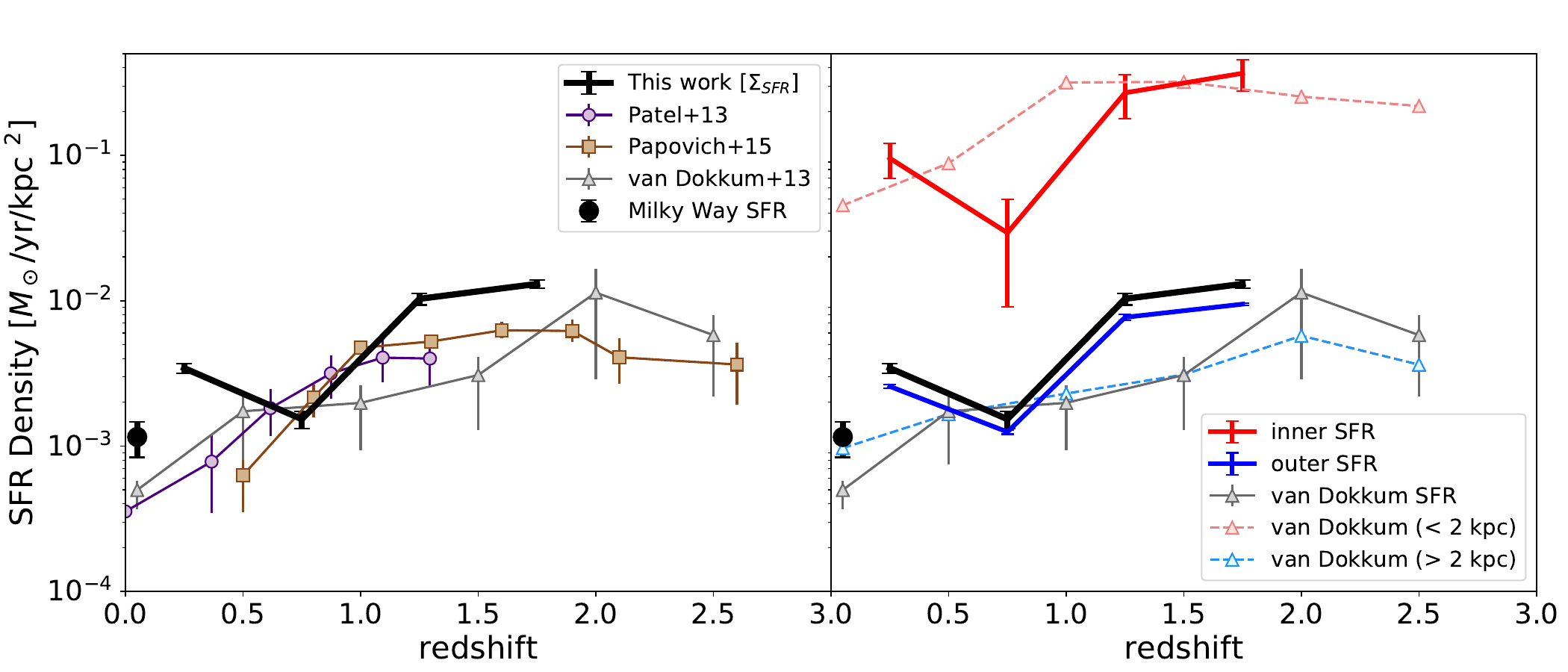}
    \caption{\emph{Left panel:} Total SFR density as a function of redshift for our results against VD13, P13, and P15. At higher redshift, from $1.5 < z < 2$, our SFR densities agree well with both P15 and P13, but the SFR density drops faster for our sample. 
    \emph{Right panel: }Total, inner, and outer SFR density as a function of redshift from our results and VD13. Their work also used UV luminosity to derive SFRs. This shows the difference that evolving comoving number density has on the SFR of the galaxies selected as progenitors. }
    \label{fig:sfr-over-time}
\end{figure*}

\subsection{Selection of MWA Progenitors and resulting difference in Mass Growth}

In Figure \ref{fig:sm-growth-compare}, we plot inner, outer, and total stellar mass density vs.~redshift, and compare them against similar plots but from VD13, P13, and P15.  The mass densities for the previous studies were obtained by taking their stellar mass as a function of redshift (inner, outer, and total) and dividing it by the area consistent with their defined limits of integration in their respective works (i.e. for VD13, the outer limit is 25 kpc, and for P13 and P15 the outer limit was defined to be 20 kpc.) Although P15 did not construct mass profiles as the other two papers did, their work still included total mass assembly for MWA progenitors. 

We see a clear difference in total stellar mass, as well as stellar mass density between P13 versus VD13 and P15. Additionally,  P13 selected MWA progenitors using the star-forming main sequence, and VD13 selecting MWA progenitors from a constant comoving density. Despite selecting our progenitors via an evolving comoving number density, our sample is less massive and more star-forming compared to VD13 and P15's analyses, and is a much closer match to P13. However, there are still differences such as the rate of stellar mass growth of the inner and outer regions.  We note that all three previous studies used IR light profiles as a proxy for stellar mass, but we are measuring stellar mass directly from the resolved 2D mass profiles.  Thus, this may account for most of the differences in mass assembly seen.

All the stellar mass density plots in Figure \ref{fig:sm-growth-compare} display signs of lockstep growth at redshifts $1 < z < 2$, but when the growth slows down at lower redshifts, it appears that the outer regions stop growing earlier than the inner regions.

\subsection{Morphological parameters as a function of time}\label{sec:results-galfit}

\begin{figure*}
    \centering
    \includegraphics[width=0.8\textwidth]{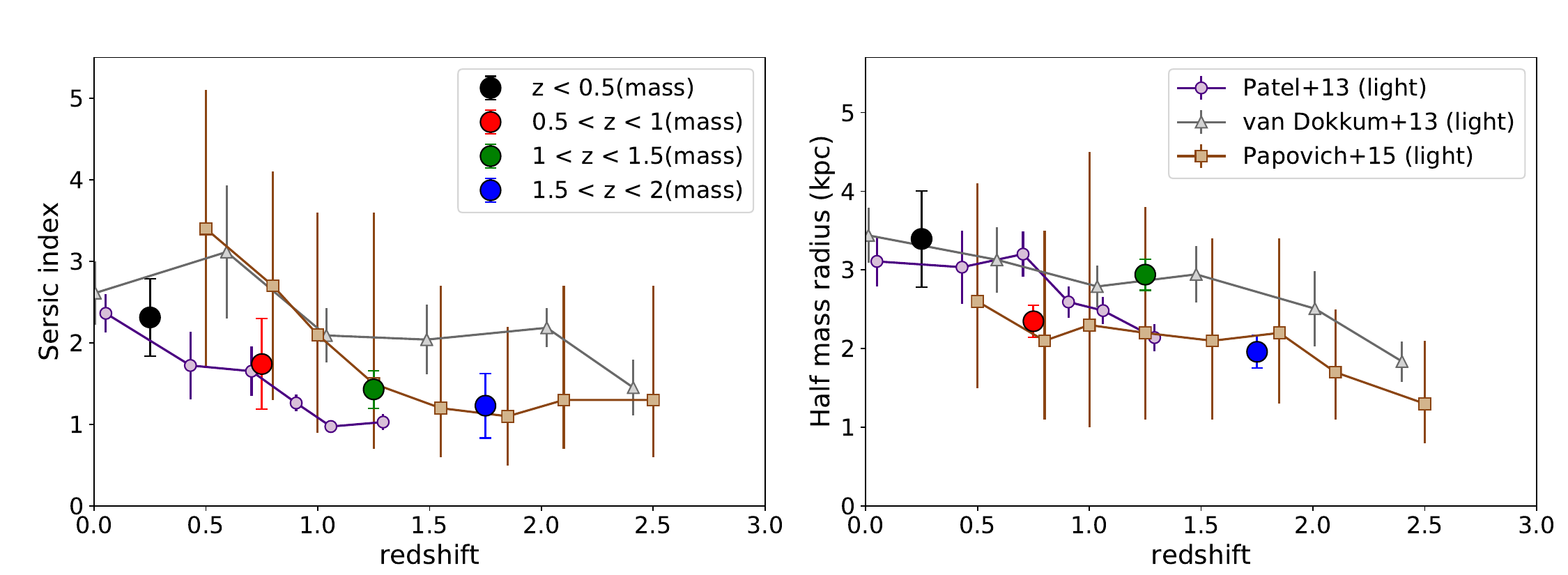}
    \caption{\emph{Left panel:} S\'ersic index vs redshift, as fit by GALFIT onto the stellar mass maps. Each point represents the median S\'ersic index for that redshift bin. Our sample has good agreement with P13's sample in S\'ersic index at similar redshifts, again showing that selecting MWA by star-formation rate and abundance matching results in similar samples. \emph{Right panel:} Half-mass radius vs redshift. Each point represents the median half-mass radius index for that redshift bin.  The average half-mass radius at each of our redshift bins are compared against the average half-light radii for P13, VD13, and P15's galaxies. }
    \label{fig:galfit-params}
\end{figure*}

In Figure \ref{fig:galfit-params}, we plot the median S\'ersic index and median half-mass radius of each redshift bin. The S\'ersic indices and effective radii were obtained by using GALFIT to fit a one-parameter S\'ersic profile on the objects' two-dimensional stellar mass maps. The error bars indicate standard error of the mean. The S\'ersic indices of the progenitor sample show a slowly increasing trend as redshift decreases at $z < 1$, but it remains constant between $1 < z < 2$. The redshift bin of $z < 0.5$ is the only epoch when the median S\'ersic index is $n > 2$. Before that epoch, from $0.5 < z < 2$, the S\'ersic index varies from 1.2 to 1.5, which suggests a disk-like morphology.

The left panel of Figure \ref{fig:galfit-params} shows that the evolution of the S\'ersic indices between this work and P15 are very close at the epoch $1 <z <2$, but diverges drastically at $z < 1$. P15's sample in fact, almost reaches a S\'ersic index of $n=4$, which describes a de Vaucoulers profile, in other words, an entirely bulge-dominated elliptical galaxy. 

 In the right panel of Figure \ref{fig:galfit-params} where the effective radii are plotted, the median half-mass radius of our sample of MWA progenitors starts at only $2.23\pm0.19$ kpc at the earliest epoch, but jumps to $3.56\pm0.21$ kpc at $1 < z < 1.5$.  The median half-mass radius is also smaller overall at $0.5 < z < 1$ than at the redshift bin immediately before and after that epoch. We do note that the quiescent fraction at that epoch is higher than at other epochs, which could contribute to a smaller median half-mass radius, since quiescent galaxies are more compact than star-forming ones. In addition, the median half-mass radii at each epoch tends to match both P13 and VD13's results but only at $z < 1$. At $1.5 < z < 2$, our results agree very well with P15, but surprisingly at $1 < z < 1.5$, it does not match the effective radii of any of the previous works. However, due to the magnitude of the error bars, the overall trend of the change in half-mass radii is consistent with the changes in half-light radii of previous MWA studies.

In general, we have found that the stellar mass profiles and growth histories are reasonably consistent with previous findings which use IR light as a proxy for stellar mass.  Given that galaxies can have color gradients (e.g., Figure \ref{fig:all-galaxies}), it is somewhat surprising how consistent these results are.  This shows that IR light tracks stellar mass quite well for MWA progenitors, even though it is not fully representative of the underlying stellar mass.  Perhaps one notable contrast is that our result in Figure \ref{fig:galfit-params} shows that the half-mass radius may grow drastically at early times, but plateaus in growth at more recent times, perhaps having to do with galaxies turning quiescent. This is somewhat different than the half-light radii, which have a more steady growth over cosmic time. The fact that the half-mass radius growth is slow, combined with the result that the S\'ersic index increasing as redshift decreases,  could imply the progenitors are growing their bulges. This is further discussed in \S\ref{disc:mass-vs-light}.

This result could also indicate lack of evolution in the half-mass radii of MWA progenitors as much as it indicates very small growth. A simulation paper on MWA progenitors in TNG50 by \cite{Hasheminia:2022} seems to find similar results (see \S \ref{disc:compare-sim} for more details).

\section{Discussion} \label{sec:discussion}
\subsection{Mass Maps vs Luminosity Maps}\label{disc:mass-vs-light}

Our analysis of MWA progenitors differs from previous work in the usage of resolved stellar mass and SFR maps instead of luminosity maps. IR luminosity is a representation of the stellar mass distribution but is not itself the mass distribution. But despite that, our results agree well with previous light profile studies. We agree particularly well with the results in P13 (as seen in Figures \ref{fig:sm-growth-compare}, \ref{fig:sfr-over-time}, and \ref{fig:galfit-params}). And our results come to the same conclusion, which is that MWA progenitors assemble their mass in inner and outer regions in lockstep since $z \sim 2$, with no preference for inside-out or outside-in growth. 

Even though P13 relied on the star-forming main sequence to select MWA progenitors, the closeness of our measurements point to a link between galaxy merger rates and the star-formation history as well as mass assembly of galaxies at a certain stellar mass in a comoving volume of space. This extends even to the closeness of the S\'ersic indices between our two studies. It may seem counter-intuitive that stellar mass profiles are quite similar to light profiles, since the Milky Way has a colour gradient with older stars near the center and younger stars further out in the disk. But looking back at the colour images of Figure \ref{fig:all-galaxies}, there are progenitors that exhibit this colour gradient at high $z$ as well as low $z$.

Our S\'ersic indices are $0.2-0.5$ greater than the S\'ersic indices of P13. Previous studies that examined the difference in S\'ersic indices between IR light profiles and stellar mass profiles include \cite{Suess:2019} and \cite{Tan:2022}, which also found $n_{mass}$ to be overall greater than $n_{IR}$ by a similar margin. This means the MWA progenitor populations are very similar between this work and P13's work, despite using a different starting mass at $z=0$. However, our half-mass radii are slightly larger. This is a notable departure from previous studies that typically found half-mass radii are smaller than half-light radii (again, both \citealt{Suess:2019} and \citealt{Tan:2022}). 

We note that differences for the two $R_e$ points in the right panel of Figure \ref{fig:galfit-params} at redshifts $0.3 < z < 1$ are minimal between this work and P13; the differences are $< 0.5$ kpc for both points. But the $R_e$ for our sample at redshift $1 < z < 1.5$ is much higher, around 1.5 kpc greater than the $R_e$  measured by P13 at the same redshift range. Since the Frontier Fields cover a smaller region of the sky than the surveys used in VD13, P13, and P15, this could be attributed to sampling bias. In addition, P13 used a stellar mass of $10^{10.5}\msun$ for MWAs at $z=0$ compared to our initial stellar mass of $10^{10.77}\msun$ for MWAs. Since our half-mass radii are the closest to their results, the difference in size could be due to the physical sizes of the chosen sample itself. Finally, our $R_e$ measurement for the redshifts $1.5 < z < 2$ lies on top of the line indicating the $R_e$ for the sample from P15, which means our $R_e$ measurements are still on the whole consistent with previous results. 

Both the MWA progenitor sample from VD13 and P15 contained more quiescent galaxies than our sample, and this can explain why their S\'ersic indices are larger at later redshifts and their half-light radii are smaller. Both of those quantities are associated with quiescent galaxies. The quiescent fractions range in P15 is 74\% to 41\% from $0 < z < 1.1$ and 29\% to 12\% from $1.1 < z < 2$. This may explain why our measurement at $1.5 < z < 2$ matches the results from P15 at the same redshift range in Figure \ref{fig:galfit-params}. It is because at that redshift there are mostly star-forming galaxies. P15 also used the abundance matching code from \cite{Behroozi:2013a} to select their MWA progenitors, so agreement at high redshift between our results and theirs may point to light and mass profiles being more similar at higher redshift, at least for star-forming galaxies. 

\subsection{Mergers and mass assembly}\label{discussion-bulge-growth}

The abundance matching code prescribes merger trees for any given $z = 0$ comoving number density or $z = 0$ halo mass, and then estimates the median cumulative number density at past epochs for any population of co-evolving galaxies \citep{Behroozi:2013a}. Due to selecting MWA progenitors via abundance matching, it raises the question of how mergers would affect the evolution of MWAs. In the case of detecting onging major mergers, there is only one in our sample, at $z\sim 1.9$. Most galaxies in our sample appear to have star-forming clumps that do not show up in the stellar mass maps (see Figure \ref{fig:galaxy-comparisons}). Despite the lack of ongoing major mergers in the sample, the assumption is still that statistically speaking, mergers still play a role in the mass assembly of MWAs on account of choosing progenitors according to an evolving co-moving number density.

If we assume the gap in total stellar mass density between our results and the results from VD13 is dominated by increase in mergers, then according to \ref{fig:sm-growth-compare}, 0.1 to 0.3 dex of the increase in total stellar mass density ($\Sigma_\star$ attributed to star-formation should instead be attributed to mergers since $z\sim 2$, as seen in the topmost panel of \ref{fig:sm-growth-compare}. For the $\Sigma_\star$ within $R < 2$kpc, this can be as high as 0.8 dex of the increase in $\Sigma_\star$, and for the region between 2 kpc $< R < $ 20 kpc, it is as high as 0.9dex, as seen in the bottom panel of \ref{fig:sm-growth-compare}.

Given that we use true stellar mass, and an evolving number density evolution, our result that the MWA progenitor S\'ersic indices are growing at a slower rate than that implied by VD13 or P15 should be a more accurate picture about bulge growth.  While all results suggests bulge strength is increasing in MWAs, ours suggest a slower evolution.  Assuming that the difference is in part driven by incorporating the merger trees into the abundance matching, this implies that mergers may affect bulge growth.  However, since the S\'ersic index growth is modest, it is clear that MW-mass galaxies that are still star-forming will not have substantial morphological transformations since $z \sim 2$.

This is reflected in both the stellar mass densities, as well as the star-formation rate densities. Figure \ref{fig:mass-assembly} shows very little difference in the stellar mass growth of the inner ($R < 2$ kpc) and outer ($R > 2$ kpc) regions compared to the total mass growth. There is a similar trend in the change in SFR versus redshift -- when the total SFR drops, both inner and outer SFRs also drop by similar rates. Since the change in SFR is not dependent on galacto-centric radius this means the newly formed stellar mass is added to both inner and outer regions, which demonstrates lockstep growth. It may seem that this contradicts the increase in S\'ersic index, but the change in S\'ersic index is rather shallow (from $n=1.23\pm0.39$ to $n=2.31\pm0.48$), and it may be that the stellar component of these galaxies compactify from dynamic friction from the gas. 

In addition, not all bulge growth is the result of mergers. Secular evolution and clump migration are other methods of bulge growth. \cite{Garrison-Kimmel:2018} find that in the FIRE-2 simulations, halo spin (halo angular momentum), is the best predictor of morphology, which mergers and secular processes (such as bars or counter-rotating disks) both greatly affect. However, clump migration in simulations has been shown to contribute very little to bulge growth despite seeing large numbers of bright clumps at high redshift \citep{Mandelker:2017,Garrison-Kimmel:2018}, with \cite{Mandelker:2017} finding that clumps only contribute $0.1 \sim 3\%$ of a galaxy's total stellar mass.  Star-forming clumps occur in the majority of $z > 1$ galaxies \citep{Elmegreen:2007}, but despite their brightness, observations do not show that clumps are any more massive than the surrounding galaxy in resolved spatial stellar mass profiles \citep{Wuyts:2012,Sok:2022}. Since the launch of JWST, galaxies with irregular features have been found to make up about 40-50\% of the universe at $3 \leq z \leq 9$ \citep{Jacobs:2023, Kartaltepe:2023} and it is clear that the progenitors of all but the most massive central galaxies in the current universe must have had irregular and clumpy morphologies. Due to the sheer amount of clumpy galaxies, their effect on mass assembly should still be investigated. Perhaps clumps play a more prominent role in galaxy evolution beyond $z = 2$.

There is a reasonable survival rate (35\%) of disks for progenitors of MW and M31 mass galaxies in the TNG50 simulation as seen in \cite{Sotillo-Ramos:2022}, showing that they do not necessarily destroy the stellar disk but only add dynamical heat to all galaxy components (i.e. in thickening the stellar disk, another process which may increase the S\'ersic index). This may also explain the lockstep growth in inner and outer regions. Depending on the properties of the merger itself, it may add stellar mass to all radii, especially if gas is distributed to the stellar disk and induces starburts. \cite{Sotillo-Ramos:2022} also find that recent major mergers of their simulated MW/M31 analogues do not appreciably deplete the gas fraction of those galaxies, allowing for further ongoing star-formation after the merger event. There is reasonable evidence from simulations that mergers play a role in bulge growth but are not dominant over secular processes, nor do they always completely transform a galaxy's morphology.

\subsection{Comparisons with MWA progenitors in simulations}\label{disc:compare-sim}
The formation of MW and M31 progenitors have been heavily studied through simulations before, and it is useful to compare the results from observations against the results from simulations. This is especially critical to answering the question of whether or not the Milky Way is an outlier in being a massive, but still disky galaxy in the local universe. Abundance matching has been applied to the IllustrisTNG simulation in \cite{Torrey:2017} to search for Milky Way and M31 progenitors in the simulation. 


\cite{Hasheminia:2022} found in their sample of MWA progenitors that the half-mass radius shows little evolution from $z \sim 2$ to the present epoch, while the S\'ersic index evolves slowly. They do find that for their galaxies, the S\'ersic index reaches upwards of $n \sim 3$ to almost $n \sim 4$, suggesting that for galaxies of Milky Way mass, they should be mostly bulge-dominated in the local universe. But we do not see this reflected in our sample from observations, where the maximum S\'ersic index at the redshift bin $0 < z < 0.5$ is only $n\sim 2.3$. However, we do reproduce one of their main results which is that the half-mass radius is roughly constant for the past 10 Gyr, despite growth in stellar mass, and change in S\'ersic index. This means there is something fundamental about the way that MWA progenitors assemble which is reflected in both simulation and observation. 

\cite{Sotillo-Ramos:2022} examined a population of MW and M31 progenitors in the TNG50 simulation, focusing specifically on the result of mergers on the diskiness of the galaxy, as well as the details of the change in morphology due to mergers. They come to two main conclusions regarding mergers: one related to bulge growth, and another related to stellar halo mass distribution. For bulge growth, they find that for their simulated MWA progenitor sample, there is no \textit{statistically significant difference} in the bulge fraction between galaxies with one or more merger events or no merger events. Secondly, they  find that for galaxies with a recent major merger in the last 5 Gyr, their bulge to halo stars ratio (with halo stars also being a proxy for disk stars) most closely match those of the MW and M31 at redshift $z=0$, \textit{but also} their bulge fractions are \textit{lower} than the rest of the sample, indicating a \textit{diskier} distribution. 

Since we obtained a lower S\'ersic index of $2.3$ at our lowest redshift bin, this implies that accounting for mergers should still result in MWA progenitors retaining a mostly disky morphology in the current universe. This is in line with results from various simulations of MWAs, which provide reasonable explanations for mass assembly, which agrees with our results. Our observational results may indicate the Milky Way is not as much of an outlier among Milky Way Analogues as was previously thought.

\section{Conclusion} \label{sec:conclusion}
We construct resolved stellar mass and SFR maps of MWA progenitors up to redshift $z \sim 2$ using resolved multi-wavelength photometry from the HFFs and examine the stellar mass assembly of Milky Way-mass galaxies in a spatially-resolved manner. We used abundance matching techniques that select progenitors with an evolving comoving number density, which is based on simulated merger histories of MW mass galaxies at $z=0$. Our main conclusions are as follows:
\begin{itemize}
    \item The stellar mass for the Milky Way Analogues increases at the same rate for both inner  ($< 2$ kpc) and outer regions ($>2$ kpc) over our redshift range, i.e. progenitors of MWAs display lockstep growth in inner and outer regions up to redshift $z\sim 2$. Our study, which focuses on resolved stellar mass profiles of galaxies agrees with previous observational results studying the Milky Way's mass assembly through light profile distributions.
    \item Generally speaking, morphological parameters derived from stellar mass profiles follow the trends seen in previous MWA progenitor studies. S\'ersic indices derived from stellar mass profiles are smaller than S\'ersic indices derived from light profiles from previous works on MWAs, with the exception of P13. However, our half-mass radii are slightly larger than the half-light radii from previous works. The difference between light-derived and mass-derived S\'ersic parameters disappears around redshift $z\sim1.5$. 
    \item We also find evidence that accounting for merger rates at different evolutionary epochs by using abundance matching in concordance with selecting by stellar mass and comoving number density results in a higher overall star-formation rate, and diskier morphology in progenitors of Milky Way Analogues. This is shown by our stellar mass density and SFR density closely matching the results of P13, which used the star-forming main sequence to select MWA progenitors at higher redshift.
    \item However, according to the SFR profiles, the SFR and the SFR density does not decrease uniformly in the inner and outer regions, and in fact the inner region decreases more rapidly than the outer regions.
    \item We find on average slightly more mass assembly in the outer 2 kpc region the MWA progenitors, especially in the epoch $0.3 < z < 0.5$, which also matches up with slightly elevated median SFR in the outer 2 kpc region at the same epoch. But it is not a large enough amount that would make our results inconsistent with lock-step growth found in previous MWA mass assembly studies.
\end{itemize}

\section{Acknowledgements} \label{sec:acknowlege}
This work is based on data and catalog products from HFF-DeepSpace, funded by the National Science Foundation and Space Telescope Science Institute (operated by the Association of Universities for Research in Astronomy, Inc., under NASA contract NAS5-26555).


\bibliography{main}{}

\begin{thebibliography}{}
\expandafter\ifx\csname natexlab\endcsname\relax\def\natexlab#1{#1}\fi
\providecommand{\url}[1]{\href{#1}{#1}}
\providecommand{\dodoi}[1]{doi:~\href{http://doi.org/#1}{\nolinkurl{#1}}}
\providecommand{\doeprint}[1]{\href{http://ascl.net/#1}{\nolinkurl{http://ascl.net/#1}}}
\providecommand{\doarXiv}[1]{\href{https://arxiv.org/abs/#1}{\nolinkurl{https://arxiv.org/abs/#1}}}

\bibitem[{{Abadi} {et~al.}(2003{\natexlab{a}}){Abadi}, {Navarro}, {Steinmetz},
  \& {Eke}}]{Abadi:2003a}
{Abadi}, M.~G., {Navarro}, J.~F., {Steinmetz}, M., \& {Eke}, V.~R.
  2003{\natexlab{a}}, \apj, 591, 499, \dodoi{10.1086/375512}

\bibitem[{{Abadi} {et~al.}(2003{\natexlab{b}}){Abadi}, {Navarro}, {Steinmetz},
  \& {Eke}}]{Abadi:2003b}
---. 2003{\natexlab{b}}, \apj, 597, 21, \dodoi{10.1086/378316}

\bibitem[{{Behroozi} {et~al.}(2013{\natexlab{a}}){Behroozi}, {Marchesini},
  {Wechsler}, {Muzzin}, {Papovich}, \& {Stefanon}}]{Behroozi:2013a}
{Behroozi}, P.~S., {Marchesini}, D., {Wechsler}, R.~H., {et~al.}
  2013{\natexlab{a}}, \apjl, 777, L10, \dodoi{10.1088/2041-8205/777/1/L10}

\bibitem[{{Behroozi} {et~al.}(2013{\natexlab{b}}){Behroozi}, {Wechsler}, \&
  {Conroy}}]{Behroozi:2013b}
{Behroozi}, P.~S., {Wechsler}, R.~H., \& {Conroy}, C. 2013{\natexlab{b}}, \apj,
  770, 57, \dodoi{10.1088/0004-637X/770/1/57}

\bibitem[{{Bertin} \& {Arnouts}(1996)}]{Bertin:1996}
{Bertin}, E., \& {Arnouts}, S. 1996, \aaps, 117, 393,
  \dodoi{10.1051/aas:1996164}

\bibitem[{{Bland-Hawthorn} \& {Gerhard}(2016)}]{Bland-Hawthorn:2016}
{Bland-Hawthorn}, J., \& {Gerhard}, O. 2016, \araa, 54, 529,
  \dodoi{10.1146/annurev-astro-081915-023441}

\bibitem[{{Bovy} \& {Rix}(2013)}]{Bovy:2013}
{Bovy}, J., \& {Rix}, H.-W. 2013, \apj, 779, 115,
  \dodoi{10.1088/0004-637X/779/2/115}

\bibitem[{{Bovy} {et~al.}(2012){Bovy}, {Rix}, \& {Hogg}}]{Bovy:2012}
{Bovy}, J., {Rix}, H.-W., \& {Hogg}, D.~W. 2012, \apj, 751, 131,
  \dodoi{10.1088/0004-637X/751/2/131}

\bibitem[{{Brada{\v{c}}} {et~al.}(2009){Brada{\v{c}}}, {Treu}, {Applegate},
  {Gonzalez}, {Clowe}, {Forman}, {Jones}, {Marshall}, {Schneider}, \&
  {Zaritsky}}]{Bradac:2009}
{Brada{\v{c}}}, M., {Treu}, T., {Applegate}, D., {et~al.} 2009, \apj, 706,
  1201, \dodoi{10.1088/0004-637X/706/2/1201}

\bibitem[{{Brammer} {et~al.}(2008){Brammer}, {van Dokkum}, \&
  {Coppi}}]{Brammer:2008}
{Brammer}, G.~B., {van Dokkum}, P.~G., \& {Coppi}, P. 2008, \apj, 686, 1503,
  \dodoi{10.1086/591786}

\bibitem[{{Bruzual} \& {Charlot}(2003)}]{Bruzual:2003}
{Bruzual}, G., \& {Charlot}, S. 2003, \mnras, 344, 1000,
  \dodoi{10.1046/j.1365-8711.2003.06897.x}

\bibitem[{{Buck} {et~al.}(2020){Buck}, {Obreja}, {Macci{\`o}}, {Minchev},
  {Dutton}, \& {Ostriker}}]{Buck:2020}
{Buck}, T., {Obreja}, A., {Macci{\`o}}, A.~V., {et~al.} 2020, \mnras, 491,
  3461, \dodoi{10.1093/mnras/stz3241}

\bibitem[{{Calzetti} {et~al.}(2000){Calzetti}, {Armus}, {Bohlin}, {Kinney},
  {Koornneef}, \& {Storchi-Bergmann}}]{Calzetti:2000}
{Calzetti}, D., {Armus}, L., {Bohlin}, R.~C., {et~al.} 2000, \apj, 533, 682,
  \dodoi{10.1086/308692}

\bibitem[{{Cappellari}(2013)}]{Cappellari:2013}
{Cappellari}, M. 2013, \apjl, 778, L2, \dodoi{10.1088/2041-8205/778/1/L2}

\bibitem[{{Cappellari} \& {Copin}(2003)}]{Cappellari:2003}
{Cappellari}, M., \& {Copin}, Y. 2003, \mnras, 342, 345,
  \dodoi{10.1046/j.1365-8711.2003.06541.x}

\bibitem[{{Chabrier}(2003)}]{Chabrier:2003}
{Chabrier}, G. 2003, \pasp, 115, 763, \dodoi{10.1086/376392}

\bibitem[{{Conroy} \& {Gunn}(2010)}]{Conroy:2010}
{Conroy}, C., \& {Gunn}, J.~E. 2010, \apj, 712, 833,
  \dodoi{10.1088/0004-637X/712/2/833}

\bibitem[{{Davidzon} {et~al.}(2017){Davidzon}, {Ilbert}, {Laigle}, {Coupon},
  {McCracken}, {Delvecchio}, {Masters}, {Capak}, {Hsieh}, {Le F{\`e}vre},
  {Tresse}, {Bethermin}, {Chang}, {Faisst}, {Le Floc'h}, {Steinhardt}, {Toft},
  {Aussel}, {Dubois}, {Hasinger}, {Salvato}, {Sanders}, {Scoville}, \&
  {Silverman}}]{Davidzon:2017}
{Davidzon}, I., {Ilbert}, O., {Laigle}, C., {et~al.} 2017, \aap, 605, A70,
  \dodoi{10.1051/0004-6361/201730419}

\bibitem[{{de Rossi} {et~al.}(2009){de Rossi}, {Tissera}, {De Lucia}, \&
  {Kauffmann}}]{deRossi:2009}
{de Rossi}, M.~E., {Tissera}, P.~B., {De Lucia}, G., \& {Kauffmann}, G. 2009,
  \mnras, 395, 210, \dodoi{10.1111/j.1365-2966.2009.14560.x}

\bibitem[{{Dekel} {et~al.}(2009){Dekel}, {Birnboim}, {Engel}, {Freundlich},
  {Goerdt}, {Mumcuoglu}, {Neistein}, {Pichon}, {Teyssier}, \&
  {Zinger}}]{Dekel:2009}
{Dekel}, A., {Birnboim}, Y., {Engel}, G., {et~al.} 2009, \nat, 457, 451,
  \dodoi{10.1038/nature07648}

\bibitem[{{Elmegreen} {et~al.}(2007){Elmegreen}, {Elmegreen}, {Ravindranath},
  \& {Coe}}]{Elmegreen:2007}
{Elmegreen}, D.~M., {Elmegreen}, B.~G., {Ravindranath}, S., \& {Coe}, D.~A.
  2007, \apj, 658, 763, \dodoi{10.1086/511667}

\bibitem[{{Freeman} \& {Bland-Hawthorn}(2002)}]{Freeman:2002}
{Freeman}, K., \& {Bland-Hawthorn}, J. 2002, \araa, 40, 487,
  \dodoi{10.1146/annurev.astro.40.060401.093840}

\bibitem[{{Freeman} {et~al.}(2013){Freeman}, {Ness}, {Wylie-de-Boer},
  {Athanassoula}, {Bland-Hawthorn}, {Asplund}, {Lewis}, {Yong}, {Lane}, {Kiss},
  \& {Ibata}}]{Freeman:2013}
{Freeman}, K., {Ness}, M., {Wylie-de-Boer}, E., {et~al.} 2013, \mnras, 428,
  3660, \dodoi{10.1093/mnras/sts305}

\bibitem[{{Garrison-Kimmel} {et~al.}(2018){Garrison-Kimmel}, {Hopkins},
  {Wetzel}, {El-Badry}, {Sanderson}, {Bullock}, {Ma}, {van de Voort}, {Hafen},
  {Faucher-Gigu{\`e}re}, {Hayward}, {Quataert}, {Kere{\v{s}}}, \&
  {Boylan-Kolchin}}]{Garrison-Kimmel:2018}
{Garrison-Kimmel}, S., {Hopkins}, P.~F., {Wetzel}, A., {et~al.} 2018, \mnras,
  481, 4133, \dodoi{10.1093/mnras/sty2513}

\bibitem[{{Gavazzi} {et~al.}(2013){Gavazzi}, {Fumagalli}, {Fossati}, {Galardo},
  {Grossetti}, {Boselli}, {Giovanelli}, \& {Haynes}}]{Gavazzi:2013}
{Gavazzi}, G., {Fumagalli}, M., {Fossati}, M., {et~al.} 2013, \aap, 553, A89,
  \dodoi{10.1051/0004-6361/201218789}

\bibitem[{{Goddard} {et~al.}(2017){Goddard}, {Thomas}, {Maraston}, {Westfall},
  {Etherington}, {Riffel}, {Mallmann}, {Zheng}, {Argudo-Fern{\'a}ndez}, {Lian},
  {Bershady}, {Bundy}, {Drory}, {Law}, {Yan}, {Wake}, {Weijmans}, {Bizyaev},
  {Brownstein}, {Lane}, {Maiolino}, {Masters}, {Merrifield}, {Nitschelm},
  {Pan}, {Roman-Lopes}, {Storchi-Bergmann}, \& {Schneider}}]{Goddard:2017}
{Goddard}, D., {Thomas}, D., {Maraston}, C., {et~al.} 2017, \mnras, 466, 4731,
  \dodoi{10.1093/mnras/stw3371}

\bibitem[{{Hasheminia} {et~al.}(2022){Hasheminia}, {Mosleh}, {Tacchella},
  {Hosseini-ShahiSavandi}, {Park}, \& {Naidu}}]{Hasheminia:2022}
{Hasheminia}, M., {Mosleh}, M., {Tacchella}, S., {et~al.} 2022, \apjl, 932,
  L23, \dodoi{10.3847/2041-8213/ac76c8}

\bibitem[{{Haywood} {et~al.}(2013){Haywood}, {Di Matteo}, {Lehnert}, {Katz}, \&
  {G{\'o}mez}}]{Haywood:2013}
{Haywood}, M., {Di Matteo}, P., {Lehnert}, M.~D., {Katz}, D., \& {G{\'o}mez},
  A. 2013, \aap, 560, A109, \dodoi{10.1051/0004-6361/201321397}

\bibitem[{{Helmi}(2020)}]{Helmi:2020}
{Helmi}, A. 2020, \araa, 58, 205, \dodoi{10.1146/annurev-astro-032620-021917}

\bibitem[{{Hill} {et~al.}(2017){Hill}, {Muzzin}, {Franx}, {Clauwens},
  {Schreiber}, {Marchesini}, {Stefanon}, {Labbe}, {Brammer}, {Caputi}, {Fynbo},
  {Milvang-Jensen}, {Skelton}, {van Dokkum}, \& {Whitaker}}]{Hill:2017a}
{Hill}, A.~R., {Muzzin}, A., {Franx}, M., {et~al.} 2017, \apj, 837, 147,
  \dodoi{10.3847/1538-4357/aa61fe}

\bibitem[{{Ibarra-Medel} {et~al.}(2016){Ibarra-Medel}, {S{\'a}nchez},
  {Avila-Reese}, {Hern{\'a}ndez-Toledo}, {Gonz{\'a}lez}, {Drory}, {Bundy},
  {Bizyaev}, {Cano-D{\'\i}az}, {Malanushenko}, {Pan}, {Roman-Lopes}, \&
  {Thomas}}]{Ibarra-Medel:2016}
{Ibarra-Medel}, H.~J., {S{\'a}nchez}, S.~F., {Avila-Reese}, V., {et~al.} 2016,
  \mnras, 463, 2799, \dodoi{10.1093/mnras/stw2126}

\bibitem[{{Jacobs} {et~al.}(2023){Jacobs}, {Glazebrook}, {Calabr{\`o}}, {Treu},
  {Nannayakkara}, {Jones}, {Merlin}, {Abraham}, {Stevens}, {Vulcani}, {Yang},
  {Bonchi}, {Boyett}, {Brada{\v{c}}}, {Castellano}, {Fontana}, {Marchesini},
  {Malkan}, {Mason}, {Morishita}, {Paris}, {Santini}, {Trenti}, \&
  {Wang}}]{Jacobs:2023}
{Jacobs}, C., {Glazebrook}, K., {Calabr{\`o}}, A., {et~al.} 2023, \apjl, 948,
  L13, \dodoi{10.3847/2041-8213/accd6d}

\bibitem[{{Karim} {et~al.}(2011){Karim}, {Schinnerer},
  {Mart{\'\i}nez-Sansigre}, {Sargent}, {van der Wel}, {Rix}, {Ilbert},
  {Smol{\v{c}}i{\'c}}, {Carilli}, {Pannella}, {Koekemoer}, {Bell}, \&
  {Salvato}}]{Karim:2011}
{Karim}, A., {Schinnerer}, E., {Mart{\'\i}nez-Sansigre}, A., {et~al.} 2011,
  \apj, 730, 61, \dodoi{10.1088/0004-637X/730/2/61}

\bibitem[{{Kartaltepe} {et~al.}(2023){Kartaltepe}, {Rose}, {Vanderhoof},
  {McGrath}, {Costantin}, {Cox}, {Yung}, {Kocevski}, {Wuyts}, {Ferguson},
  {Bagley}, {Finkelstein}, {Amor{\'\i}n}, {Andrews}, {Haro}, {Backhaus},
  {Behroozi}, {Bisigello}, {Calabr{\`o}}, {Casey}, {Coogan}, {Cooper},
  {Croton}, {de la Vega}, {Dickinson}, {Fontana}, {Franco}, {Grazian},
  {Grogin}, {Hathi}, {Holwerda}, {Huertas-Company}, {Iyer}, {Jogee}, {Jung},
  {Kewley}, {Kirkpatrick}, {Koekemoer}, {Liu}, {Lotz}, {Lucas}, {Newman},
  {Pacifici}, {Pandya}, {Papovich}, {Pentericci}, {P{\'e}rez-Gonz{\'a}lez},
  {Petersen}, {Pirzkal}, {Rafelski}, {Ravindranath}, {Simons}, {Snyder},
  {Somerville}, {Stanway}, {Straughn}, {Tacchella}, {Trump}, {Vega-Ferrero},
  {Wilkins}, {Yang}, \& {Zavala}}]{Kartaltepe:2023}
{Kartaltepe}, J.~S., {Rose}, C., {Vanderhoof}, B.~N., {et~al.} 2023, \apjl,
  946, L15, \dodoi{10.3847/2041-8213/acad01}

\bibitem[{{Kennicutt} \& {Evans}(2012)}]{Kennicutt:2012}
{Kennicutt}, R.~C., \& {Evans}, N.~J. 2012, \araa, 50, 531,
  \dodoi{10.1146/annurev-astro-081811-125610}

\bibitem[{{Kravtsov} {et~al.}(2018){Kravtsov}, {Vikhlinin}, \&
  {Meshcheryakov}}]{Kravtsov:2018}
{Kravtsov}, A.~V., {Vikhlinin}, A.~A., \& {Meshcheryakov}, A.~V. 2018,
  Astronomy Letters, 44, 8, \dodoi{10.1134/S1063773717120015}

\bibitem[{{Kriek} {et~al.}(2009){Kriek}, {van Dokkum}, {Labb{\'e}}, {Franx},
  {Illingworth}, {Marchesini}, \& {Quadri}}]{Kriek:2009}
{Kriek}, M., {van Dokkum}, P.~G., {Labb{\'e}}, I., {et~al.} 2009, \apj, 700,
  221, \dodoi{10.1088/0004-637X/700/1/221}

\bibitem[{{Leitner}(2012)}]{Leitner:2012}
{Leitner}, S.~N. 2012, \apj, 745, 149, \dodoi{10.1088/0004-637X/745/2/149}

\bibitem[{{Leja} {et~al.}(2013){Leja}, {van Dokkum}, \& {Franx}}]{Leja:2013}
{Leja}, J., {van Dokkum}, P., \& {Franx}, M. 2013, \apj, 766, 33,
  \dodoi{10.1088/0004-637X/766/1/33}

\bibitem[{{Licquia} {et~al.}(2015){Licquia}, {Newman}, \&
  {Brinchmann}}]{Licquia:2015b}
{Licquia}, T.~C., {Newman}, J.~A., \& {Brinchmann}, J. 2015, \apj, 809, 96,
  \dodoi{10.1088/0004-637X/809/1/96}

\bibitem[{{Lin} {et~al.}(2013){Lin}, {Brodwin}, {Gonzalez}, {Bode},
  {Eisenhardt}, {Stanford}, \& {Vikhlinin}}]{Lin:2013}
{Lin}, Y.-T., {Brodwin}, M., {Gonzalez}, A.~H., {et~al.} 2013, \apj, 771, 61,
  \dodoi{10.1088/0004-637X/771/1/61}

\bibitem[{{Lotz} {et~al.}(2017){Lotz}, {Koekemoer}, {Coe}, {Grogin}, {Capak},
  {Mack}, {Anderson}, {Avila}, {Barker}, {Borncamp}, {Brammer}, {Durbin},
  {Gunning}, {Hilbert}, {Jenkner}, {Khandrika}, {Levay}, {Lucas}, {MacKenty},
  {Ogaz}, {Porterfield}, {Reid}, {Robberto}, {Royle}, {Smith},
  {Storrie-Lombardi}, {Sunnquist}, {Surace}, {Taylor}, {Williams}, {Bullock},
  {Dickinson}, {Finkelstein}, {Natarajan}, {Richard}, {Robertson}, {Tumlinson},
  {Zitrin}, {Flanagan}, {Sembach}, {Soifer}, \& {Mountain}}]{Lotz:2017}
{Lotz}, J.~M., {Koekemoer}, A., {Coe}, D., {et~al.} 2017, \apj, 837, 97,
  \dodoi{10.3847/1538-4357/837/1/97}

\bibitem[{{Mandelker} {et~al.}(2017){Mandelker}, {Dekel}, {Ceverino}, {DeGraf},
  {Guo}, \& {Primack}}]{Mandelker:2017}
{Mandelker}, N., {Dekel}, A., {Ceverino}, D., {et~al.} 2017, \mnras, 464, 635,
  \dodoi{10.1093/mnras/stw2358}

\bibitem[{{Marchesini} {et~al.}(2009){Marchesini}, {van Dokkum}, {F{\"o}rster
  Schreiber}, {Franx}, {Labb{\'e}}, \& {Wuyts}}]{Marchesini:2009}
{Marchesini}, D., {van Dokkum}, P.~G., {F{\"o}rster Schreiber}, N.~M., {et~al.}
  2009, \apj, 701, 1765, \dodoi{10.1088/0004-637X/701/2/1765}

\bibitem[{{Marchesini} {et~al.}(2014){Marchesini}, {Muzzin}, {Stefanon},
  {Franx}, {Brammer}, {Marsan}, {Vulcani}, {Fynbo}, {Milvang-Jensen}, {Dunlop},
  \& {Buitrago}}]{Marchesini:2014}
{Marchesini}, D., {Muzzin}, A., {Stefanon}, M., {et~al.} 2014, \apj, 794, 65,
  \dodoi{10.1088/0004-637X/794/1/65}

\bibitem[{{Martig} {et~al.}(2009){Martig}, {Bournaud}, {Teyssier}, \&
  {Dekel}}]{Martig:2009}
{Martig}, M., {Bournaud}, F., {Teyssier}, R., \& {Dekel}, A. 2009, \apj, 707,
  250, \dodoi{10.1088/0004-637X/707/1/250}

\bibitem[{{McCracken} {et~al.}(2012){McCracken}, {Milvang-Jensen}, {Dunlop},
  {Franx}, {Fynbo}, {Le F{\`e}vre}, {Holt}, {Caputi}, {Goranova}, {Buitrago},
  {Emerson}, {Freudling}, {Hudelot}, {L{\'o}pez-Sanjuan}, {Magnard}, {Mellier},
  {M{\o}ller}, {Nilsson}, {Sutherland}, {Tasca}, \& {Zabl}}]{McCracken:2012}
{McCracken}, H.~J., {Milvang-Jensen}, B., {Dunlop}, J., {et~al.} 2012, \aap,
  544, A156, \dodoi{10.1051/0004-6361/201219507}

\bibitem[{{Minniti} {et~al.}(2011){Minniti}, {Saito}, {Alonso-Garc{\'\i}a},
  {Lucas}, \& {Hempel}}]{Minniti:2011}
{Minniti}, D., {Saito}, R.~K., {Alonso-Garc{\'\i}a}, J., {Lucas}, P.~W., \&
  {Hempel}, M. 2011, \apjl, 733, L43, \dodoi{10.1088/2041-8205/733/2/L43}

\bibitem[{{Mosleh} {et~al.}(2017){Mosleh}, {Tacchella}, {Renzini}, {Carollo},
  {Molaeinezhad}, {Onodera}, {Khosroshahi}, \& {Lilly}}]{Mosleh:2017}
{Mosleh}, M., {Tacchella}, S., {Renzini}, A., {et~al.} 2017, \apj, 837, 2,
  \dodoi{10.3847/1538-4357/aa5f14}

\bibitem[{{Moster} {et~al.}(2013){Moster}, {Naab}, \& {White}}]{Moster:2013}
{Moster}, B.~P., {Naab}, T., \& {White}, S. D.~M. 2013, \mnras, 428, 3121,
  \dodoi{10.1093/mnras/sts261}

\bibitem[{{Mu{\~n}oz-Mateos} {et~al.}(2007){Mu{\~n}oz-Mateos}, {Gil de Paz},
  {Boissier}, {Zamorano}, {Jarrett}, {Gallego}, \&
  {Madore}}]{Munoz-Mateos:2007}
{Mu{\~n}oz-Mateos}, J.~C., {Gil de Paz}, A., {Boissier}, S., {et~al.} 2007,
  \apj, 658, 1006, \dodoi{10.1086/511812}

\bibitem[{{Mundy} {et~al.}(2015){Mundy}, {Conselice}, \&
  {Ownsworth}}]{Mundy:2015}
{Mundy}, C.~J., {Conselice}, C.~J., \& {Ownsworth}, J.~R. 2015, \mnras, 450,
  3696, \dodoi{10.1093/mnras/stv860}

\bibitem[{{Mutch} {et~al.}(2011){Mutch}, {Croton}, \& {Poole}}]{Mutch:2011}
{Mutch}, S.~J., {Croton}, D.~J., \& {Poole}, G.~B. 2011, \apj, 736, 84,
  \dodoi{10.1088/0004-637X/736/2/84}

\bibitem[{{Muzzin} {et~al.}(2013{\natexlab{a}}){Muzzin}, {Marchesini},
  {Stefanon}, {Franx}, {McCracken}, {Milvang-Jensen}, {Dunlop}, {Fynbo},
  {Brammer}, {Labb{\'e}}, \& {van Dokkum}}]{Muzzin:2013b}
{Muzzin}, A., {Marchesini}, D., {Stefanon}, M., {et~al.} 2013{\natexlab{a}},
  \apj, 777, 18, \dodoi{10.1088/0004-637X/777/1/18}

\bibitem[{{Muzzin} {et~al.}(2013{\natexlab{b}}){Muzzin}, {Marchesini},
  {Stefanon}, {Franx}, {Milvang-Jensen}, {Dunlop}, {Fynbo}, {Brammer},
  {Labb{\'e}}, \& {van Dokkum}}]{Muzzin:2013a}
---. 2013{\natexlab{b}}, \apjs, 206, 8, \dodoi{10.1088/0067-0049/206/1/8}

\bibitem[{{O{\~n}orbe} {et~al.}(2015){O{\~n}orbe}, {Boylan-Kolchin}, {Bullock},
  {Hopkins}, {Kere{\v{s}}}, {Faucher-Gigu{\`e}re}, {Quataert}, \&
  {Murray}}]{Onorbe:2015}
{O{\~n}orbe}, J., {Boylan-Kolchin}, M., {Bullock}, J.~S., {et~al.} 2015,
  \mnras, 454, 2092, \dodoi{10.1093/mnras/stv2072}

\bibitem[{{Ownsworth} {et~al.}(2014){Ownsworth}, {Conselice}, {Mortlock},
  {Hartley}, {Almaini}, {Duncan}, \& {Mundy}}]{Ownsworth:2014}
{Ownsworth}, J.~R., {Conselice}, C.~J., {Mortlock}, A., {et~al.} 2014, \mnras,
  445, 2198, \dodoi{10.1093/mnras/stu1802}

\bibitem[{{Ownsworth} {et~al.}(2016){Ownsworth}, {Conselice}, {Mundy},
  {Mortlock}, {Hartley}, {Duncan}, \& {Almaini}}]{Ownsworth:2016}
{Ownsworth}, J.~R., {Conselice}, C.~J., {Mundy}, C.~J., {et~al.} 2016, \mnras,
  461, 1112, \dodoi{10.1093/mnras/stw1207}

\bibitem[{{Papovich} {et~al.}(2015){Papovich}, {Labb{\'e}}, {Quadri}, {Tilvi},
  {Behroozi}, {Bell}, {Glazebrook}, {Spitler}, {Straatman}, {Tran}, {Cowley},
  {Dav{\'e}}, {Dekel}, {Dickinson}, {Ferguson}, {Finkelstein}, {Gawiser},
  {Inami}, {Faber}, {Kacprzak}, {Kawinwanichakij}, {Kocevski}, {Koekemoer},
  {Koo}, {Kurczynski}, {Lotz}, {Lu}, {Lucas}, {McIntosh}, {Mehrtens},
  {Mobasher}, {Monson}, {Morrison}, {Nanayakkara}, {Persson}, {Salmon},
  {Simons}, {Tomczak}, {van Dokkum}, {Weiner}, \& {Willner}}]{Papovich:2015}
{Papovich}, C., {Labb{\'e}}, I., {Quadri}, R., {et~al.} 2015, \apj, 803, 26,
  \dodoi{10.1088/0004-637X/803/1/26}

\bibitem[{{Patel} {et~al.}(2013){Patel}, {Fumagalli}, {Franx}, {van Dokkum},
  {van der Wel}, {Leja}, {Labb{\'e}}, {Brammer}, {Skelton}, {Momcheva},
  {Whitaker}, {Lundgren}, {Muzzin}, {Quadri}, {Nelson}, {Wake}, \&
  {Rix}}]{Patel:2013}
{Patel}, S.~G., {Fumagalli}, M., {Franx}, M., {et~al.} 2013, \apj, 778, 115,
  \dodoi{10.1088/0004-637X/778/2/115}

\bibitem[{{Peng} {et~al.}(2002){Peng}, {Ho}, {Impey}, \& {Rix}}]{Peng:2002}
{Peng}, C.~Y., {Ho}, L.~C., {Impey}, C.~D., \& {Rix}, H.-W. 2002, \aj, 124,
  266, \dodoi{10.1086/340952}

\bibitem[{{Rix} \& {Bovy}(2013)}]{Rix:2013}
{Rix}, H.-W., \& {Bovy}, J. 2013, A\&A Rev., 21, 61,
  \dodoi{10.1007/s00159-013-0061-8}

\bibitem[{{Robin} {et~al.}(1992){Robin}, {Creze}, \& {Mohan}}]{Robin:1992}
{Robin}, A.~C., {Creze}, M., \& {Mohan}, V. 1992, \apjl, 400, L25,
  \dodoi{10.1086/186640}

\bibitem[{{Ruphy} {et~al.}(1996){Ruphy}, {Robin}, {Epchtein}, {Copet},
  {Bertin}, {Fouque}, \& {Guglielmo}}]{Ruphy:1996}
{Ruphy}, S., {Robin}, A.~C., {Epchtein}, N., {et~al.} 1996, \aap, 313, L21

\bibitem[{{Schaefer} {et~al.}(2017){Schaefer}, {Croom}, {Allen}, {Brough},
  {Medling}, {Ho}, {Scott}, {Richards}, {Pracy}, {Gunawardhana}, {Norberg},
  {Alpaslan}, {Bauer}, {Bekki}, {Bland-Hawthorn}, {Bloom}, {Bryant}, {Couch},
  {Driver}, {Fogarty}, {Foster}, {Goldstein}, {Green}, {Hopkins},
  {Konstantopoulos}, {Lawrence}, {L{\'o}pez-S{\'a}nchez}, {Lorente}, {Owers},
  {Sharp}, {Sweet}, {Taylor}, {van de Sande}, {Walcher}, \&
  {Wong}}]{Schaefer:2017}
{Schaefer}, A.~L., {Croom}, S.~M., {Allen}, J.~T., {et~al.} 2017, \mnras, 464,
  121, \dodoi{10.1093/mnras/stw2289}

\bibitem[{{Schreiber} {et~al.}(2015){Schreiber}, {Pannella}, {Elbaz},
  {B{\'e}thermin}, {Inami}, {Dickinson}, {Magnelli}, {Wang}, {Aussel}, {Daddi},
  {Juneau}, {Shu}, {Sargent}, {Buat}, {Faber}, {Ferguson}, {Giavalisco},
  {Koekemoer}, {Magdis}, {Morrison}, {Papovich}, {Santini}, \&
  {Scott}}]{Schreiber:2015}
{Schreiber}, C., {Pannella}, M., {Elbaz}, D., {et~al.} 2015, \aap, 575, A74,
  \dodoi{10.1051/0004-6361/201425017}

\bibitem[{{Shipley} {et~al.}(2018){Shipley}, {Lange-Vagle}, {Marchesini},
  {Brammer}, {Ferrarese}, {Stefanon}, {Kado-Fong}, {Whitaker}, {Oesch},
  {Feinstein}, {Labb{\'e}}, {Lundgren}, {Martis}, {Muzzin}, {Nedkova},
  {Skelton}, \& {van der Wel}}]{Shipley:2018}
{Shipley}, H.~V., {Lange-Vagle}, D., {Marchesini}, D., {et~al.} 2018, \apjs,
  235, 14, \dodoi{10.3847/1538-4365/aaacce}

\bibitem[{{Sok} {et~al.}(2022){Sok}, {Muzzin}, {Jablonka}, {Marsan}, {Tan},
  {Alcorn}, {Marchesini}, \& {Stefanon}}]{Sok:2022}
{Sok}, V., {Muzzin}, A., {Jablonka}, P., {et~al.} 2022, \apj, 924, 7,
  \dodoi{10.3847/1538-4357/ac2f40}

\bibitem[{{Sotillo-Ramos} {et~al.}(2022){Sotillo-Ramos}, {Pillepich},
  {Donnari}, {Nelson}, {Eisert}, {Rodriguez-Gomez}, {Joshi}, {Vogelsberger}, \&
  {Hernquist}}]{Sotillo-Ramos:2022}
{Sotillo-Ramos}, D., {Pillepich}, A., {Donnari}, M., {et~al.} 2022, \mnras,
  516, 5404, \dodoi{10.1093/mnras/stac2586}

\bibitem[{{Suess} {et~al.}(2019){Suess}, {Kriek}, {Price}, \&
  {Barro}}]{Suess:2019}
{Suess}, K.~A., {Kriek}, M., {Price}, S.~H., \& {Barro}, G. 2019, \apj, 877,
  103, \dodoi{10.3847/1538-4357/ab1bda}

\bibitem[{{Tan} {et~al.}(2022){Tan}, {Muzzin}, {Marsan}, {Sok}, {Alcorn},
  {Matharu}, {Shipley}, {Marchesini}, {Nedkova}, {Martis}, {van der Wel}, \&
  {Whitaker}}]{Tan:2022}
{Tan}, V. Y.~Y., {Muzzin}, A., {Marsan}, Z.~C., {et~al.} 2022, \apj, 933, 30,
  \dodoi{10.3847/1538-4357/ac7051}

\bibitem[{{Torrey} {et~al.}(2017){Torrey}, {Wellons}, {Ma}, {Hopkins}, \&
  {Vogelsberger}}]{Torrey:2017}
{Torrey}, P., {Wellons}, S., {Ma}, C.-P., {Hopkins}, P.~F., \& {Vogelsberger},
  M. 2017, \mnras, 467, 4872, \dodoi{10.1093/mnras/stx370}

\bibitem[{{van den Bosch}(1998)}]{vandenBosch:1998}
{van den Bosch}, F.~C. 1998, \apj, 507, 601, \dodoi{10.1086/306354}

\bibitem[{{van Dokkum} {et~al.}(2013){van Dokkum}, {Leja}, {Nelson}, {Patel},
  {Skelton}, {Momcheva}, {Brammer}, {Whitaker}, {Lundgren}, {Fumagalli},
  {Conroy}, {F{\"o}rster Schreiber}, {Franx}, {Kriek}, {Labb{\'e}},
  {Marchesini}, {Rix}, {van der Wel}, \& {Wuyts}}]{vanDokkum:2013}
{van Dokkum}, P.~G., {Leja}, J., {Nelson}, E.~J., {et~al.} 2013, \apjl, 771,
  L35, \dodoi{10.1088/2041-8205/771/2/L35}

\bibitem[{{Wang} {et~al.}(2011){Wang}, {Kauffmann}, {Overzier}, {Catinella},
  {Schiminovich}, {Heckman}, {Moran}, {Haynes}, {Giovanelli}, \&
  {Kong}}]{Wang:2011}
{Wang}, J., {Kauffmann}, G., {Overzier}, R., {et~al.} 2011, \mnras, 412, 1081,
  \dodoi{10.1111/j.1365-2966.2010.17962.x}

\bibitem[{{Wegg} {et~al.}(2015){Wegg}, {Gerhard}, \& {Portail}}]{Wegg:2015}
{Wegg}, C., {Gerhard}, O., \& {Portail}, M. 2015, \mnras, 450, 4050,
  \dodoi{10.1093/mnras/stv745}

\bibitem[{{Wuyts} {et~al.}(2012){Wuyts}, {F{\"o}rster Schreiber}, {Genzel},
  {Guo}, {Barro}, {Bell}, {Dekel}, {Faber}, {Ferguson}, {Giavalisco}, {Grogin},
  {Hathi}, {Huang}, {Kocevski}, {Koekemoer}, {Koo}, {Lotz}, {Lutz}, {McGrath},
  {Newman}, {Rosario}, {Saintonge}, {Tacconi}, {Weiner}, \& {van der
  Wel}}]{Wuyts:2012}
{Wuyts}, S., {F{\"o}rster Schreiber}, N.~M., {Genzel}, R., {et~al.} 2012, \apj,
  753, 114, \dodoi{10.1088/0004-637X/753/2/114}

\bibitem[{{Yang} {et~al.}(2012){Yang}, {Mo}, {van den Bosch}, {Zhang}, \&
  {Han}}]{Yang:2012}
{Yang}, X., {Mo}, H.~J., {van den Bosch}, F.~C., {Zhang}, Y., \& {Han}, J.
  2012, \apj, 752, 41, \dodoi{10.1088/0004-637X/752/1/41}

\end{thebibliography}
\bibliographystyle{aasjournal}



\end{document}